\documentclass[fleqn,usenatbib]{mnras}

\usepackage{newtxtext,newtxmath}

\usepackage[T1]{fontenc}

\DeclareRobustCommand{\VAN}[3]{#2}
\let\VANthebibliography\thebibliography
\def\thebibliography{\DeclareRobustCommand{\VAN}[3]{##3}\VANthebibliography}

\usepackage[dvipdfmx]{graphicx}	
\usepackage{amsmath}	
\usepackage{bm}
\usepackage{here}
\usepackage{newtxtext,newtxmath}

\newcommand{\beq}{\begin{equation}}
\newcommand{\beqa}{\begin{eqnarray}}
\newcommand{\eeq}{\end{equation}}
\newcommand{\eeqa}{\end{eqnarray}}

\newcommand{\pmem}{p_{\rm mem}}

\newcommand{\msun}{h^{-1}{\rm M}_{\odot}}

\newcommand{\mpch}{h^{-1}{\rm Mpc}}

\newcommand{\simgt}{\lower.5ex\hbox{$\; \buildrel > \over \sim \;$}}
\newcommand{\simlt}{\lower.5ex\hbox{$\; \buildrel < \over \sim \;$}}

\title[Anisotropic boost of galaxy clusters]{Observational constraints of an anisotropic boost due to the projection effects using redMaPPer clusters}

\author[T.~Sunayama]{Tomomi Sunayama\thanks{E-mail: sunayama.tomomi.f2@a.mail.nagoya-u.ac.jp}$^{1}$
\newauthor
\\
$^{1}$ Kobayashi-Maskawa Institute for the Origin of Particles and the Universe (KMI), Nagoya University, Nagoya, 464-8602, Japan\\
}
\date{Accepted XXX. Received YYY; in original form ZZZ}

\pubyear{2022}

\begin{document}
\label{firstpage}
\pagerange{\pageref{firstpage}--\pageref{lastpage}}
\maketitle

\begin{abstract}
Optical clusters identified from red-sequence galaxies suffer from projection effects, where 
interloper galaxies along the line-of-sight to a cluster are mistaken as genuine members of the cluster.
In the previous study \citep{Sunayama_etal2020}, we found that the projection effects cause the boost on the amplitudes of clustering and lensing on large scale compared to the expected amplitudes in the absence of any projection effects.
These boosts are caused by preferential selections of filamentary structure aligned to the line-of-sight due to distance uncertainties in photometric surveys.
We model the projection effects with two simple assumptions and develop a novel method to quantify the size of the boost using cluster-galaxy cross-correlation functions.
We validate our method using mock cluster catalogs built from cosmological N-body simulations and find that we can obtain unbiased constraints on the boost parameter with our model. 
We then apply our analysis on the SDSS redMaPPer clusters and find that the size of the boost is roughly $20\%$ for all the richness bins except the cluster sample with the richness bin $\lambda \in [30,40]$.
This is the first study to constrain the boost parameter independent from cluster cosmology studies and provides a self-consistency test for the projection effects.
\end{abstract}




\section{Introduction}
Galaxy clusters are the most massive gravitationally self-bound objects in the Universe. These clusters form at the rare high peaks of the initial density field and the abundance of clusters and its time evolution are sensitive to the growth of structure in the Universe. Hence, clusters have been used to constrain cosmological parameters \citep{Whieetal:93,Haiman:2001,Vikhlininetal:09,Rozoetal:10,TakadaBridle:07,OguriTakada:2011} \citep[also see][for a review]{Weinberg:2013}. 
Many ongoing and future galaxy surveys, such as the Hyper Suprime-Cam (HSC) survey \citep{HSCOverview:17}, the Dark Energy Survey\footnote{\url{ https://www.darkenergysurvey.org}} (DES) \citep{DES2005}, the Kilo Degree Survey\footnote{\url{http://kids.strw.leidenuniv.nl/}} (KiDS) \citep{KiDs2015}, the Rubin Observatory Legacy Survey of Space and Time\footnote{\url{https://www.lsst.org}} (LSST) \citep{LSST2009}, \textit{Euclid}\footnote{\url{ https://sci.esa.int/web/euclid}} \citep{euclid2018}, and the Nancy Grace Roman Telescope\footnote{\url{https://wfirst.gsfc.nasa.gov}} \citep{WFIRST2019}, will provide unprecedented numbers of clusters and enable us to carry out cluster cosmology analyses with great precision if all the systematic effects are under control. 
In particular, these optically identified clusters from the photometric surveys are known to be susceptible to the systematic effects due to the photometric redshift uncertainties of galaxies.

One of the main systematic effects for optically identified clusters is so-called “projection effects” that interloper galaxies along the line-of-sight (LOS) to a cluster are mistakenly identified as members of the cluster. 
The projection effects alter the mass-observable relation such that the observable for the optical clusters, which is the weighted sum of member galaxies (referred to “richness”), is boosted concerning its halo mass \citep[e.g.][]{costanzietalprojection}. 
In addition to the alternation of the mass-richness relation, \cite{Sunayama_etal2020} found that projection effects boost the amplitude of cluster lensing and clustering signals on large scales due to the preferential identification of filaments aligned with the LOS direction as a cluster. 
This results in the anisotropic distribution of matter around the optical clusters.
The predicted size of this anisotropic boost from \cite{Sunayama_etal2020} is roughly $20-30\%$. 
These anisotropic boosts have been parameterized in a few cluster cosmology analysis. \cite{To_Krause2021} modeled the boost in their combined cosmology analysis with clusters and galaxies and obtained a similar size of the boost as the study by \cite{Sunayama_etal2020}. 
\cite{Park_etal2021} also employed this boost model in their cluster cosmology analyses and applied their full forward modeling method to the red-sequence Matchedfilter Probabilistic Percolation (redMaPPer) cluster catalog \citep{Rykoff_etal2014}, constructed from the Sloan Digital Sky Survey (SDSS) DR8 data \citep{Aihara:2011}. 
While the constrained value for the boost parameter in \cite{Park_etal2021} was consistent with \cite{To_Krause2021}, the result for the cosmological parameters favored low $\Omega_m$ and high $\sigma_8$. This questions how the boost exactly manifests in the real data observables, i.e., in the measured lensing and clustering signals.
    
Even though the boost of the lensing and clustering amplitudes has been modeled and constrained in these cosmology analyses, no study finds a direct evidence of this effect. 
The goal of this study is to quantitatively constrain the size of this anisotropic boost in the amplitude of clustering and lensing signals seen in \cite{Sunayama_etal2020}. For this, we develop a novel method to quantify the boost caused by the projection effects with two simple assumptions using cluster-galaxy cross-correlation functions. 
The cross-correlation with spectroscopic galaxies provides a way to evaluate the anisotropic structure around the optical clusters.
We first validate our method using mock cluster catalogs built from cosmological N-body simulations, and measure the anisotropic boost parameter in the SDSS redMaPPer clusters using this method. 
This is the first study to directly constrain the boost parameter using cluster-galaxy cross-correlation functions and provide a self-consistency test for the projection effects.

This paper is organized as follows: in Section 2 we describe the details of the mock cluster catalog as well as the galaxy cluster sample. In Section 3 we explain our method to quantify the boost using the cluster-galaxy cross-correlation functions. In Section 4, we first validate our method using the mock cluster catalog and apply it to the SDSS redMaPPer clusters. We summarize our work and discuss its implication in Section 5.

\section{Data}
\label{sec:data}
In this section, we describe the details of the simulations as well as how we build the galaxy and cluster mock catalogs. In principle, we are following these procedures from \cite{Sunayama_etal2020}.

\subsection{$N$-body Simulations and Halo Catalogs}
\label{sec:sim:nbody}

We use the $N$-body simulations and halo catalogs from \cite{darkemu}. These $N$-body simulations were performed with $2048^3$ particles in a comoving cubic box with side lengths of $1h^{-1}{\rm Gpc}$, assuming the best-fit flat $\Lambda$CDM model\footnote{$\{\omega_{\rm b},\omega_{\rm c},\Omega_{\rm \Lambda},\ln(10^{10}A_{\rm s}),n_{\rm s}\} = \{0.02225,0.1198,0.6844,3.094,0.9645\}$} from \textit{Planck} Data Release 2 \citep{Planck:2015}. The initial displacement vector and the initial velocity of each $N$-body particle was set by second-order Lagrangian perturbation theory \citep{scoccimarro98,crocce06a,crocce06b,nishimichi09} with an input linear matter power spectrum computed from the publicly available Boltzmann code {\tt CAMB} \citep{camb}, and the subsequent time evolution of the particle distribution was simulated using the parallel Tree-Particle Mesh code {\tt Gadget2} \citep{Springel:2005}. 
The {\it Planck} model has $\Omega_{\rm m}=0.3156$
(the present-day matter density parameter), $\sigma_8=0.831$ (the present-day RMS linear mass density fluctuations within a top-hat sphere of radius $8~h^{-1}{\rm Mpc}$) and $h=0.672$ for the Hubble parameter. The particle mass is $1.02\times 10^{10}~h^{-1}M_\odot$. 

To generate halo catalogs, we first take simulation snapshots at redshift $z=0.251$ -- chosen to be close to the mean redshift of SDSS redMaPPer clusters -- and identify halos using the Friends-of-Friends (FoF) halo finder \texttt{Rockstar} developed in \cite{Behroozi:2013} \citep[also see][for details]{darkemu}. We use the ``200m'' halo definition, defining halo masses as $M\equiv M_{\rm 200m}=(4\pi/3)(R_{\rm 200m})^3(200\bar{\rho}_{\rm m0})$ where $R_{\rm 200m}$ is the spherical halo boundary radius within which the mean mass density is $200\times \bar{\rho}_{\rm m0}$, where we use the present-day mean mass density $\bar{\rho}_{\rm m0}$ using comoving coordinates. Note that we use comoving length units for $R_{\rm 200m}$.

Our definition of halo mass includes all particles within the radius $R_{\rm 200m}$ from the halo center, i.e. includes particles even if they are not gravitationally bound to the halo. We only keep halos with masses above $10^{12}\,h^{-1}M_\odot$ in the final halo catalog used in this paper. The ``minimum halo'' at
$M=10^{12}~h^{-1}M_\odot$ consists of 100 $N$-body particles.

\subsection{Mock Catalogs of Red-Sequence Galaxies/LOWZ Galaxies}
\label{sec:sim:mocks_galaxy}
We construct two types of mock catalogs from the $N$-body simulations discussed in Section~\ref{sec:sim:nbody}. One is mock red-sequence galaxies which represent the SDSS DR8 photometric galaxy catalog \citep{Aihara:2011} to build a cluster mock catalog, and the other is mock bright galaxies whose number in a halo and spatial distribution are similar to the BOSS LOWZ galaxies \citep{parejko_etal2012}.
To construct these mock catalogs, we use the halo occupation distribution (HOD) formulation \citep{1998ApJ...494....1J,seljak:2000uq,peacock:2000qy,2005ApJ...633..791Z} to populate mock galaxies in halos. Our HOD model gives the expected numbers of central and satellite galaxies, $N_{\rm cen}(M)$ and $N_{\rm sat}(M)$, as a function of halo mass $M$:
\begin{align}
N_{\rm cen}(M)=
\frac{1}{2}\left[
1+{\rm erf}\left(\frac{\log M-\log M_{\rm cut}}{\sigma_{\log M}}\right)
\right]
\end{align}
and
\begin{align}
N_{\rm sat}(M)=N_{\rm cen}(M)\left(\frac{M-\kappa M_{\rm cut}}{M_1}\right)^\alpha ,
\label{eq:HOD_sat}
\end{align}
where $M_{\rm cut}, M_1, \sigma_{\log M}, \kappa,$ and $\alpha$ are model parameters. 

With the HOD prescription in hand, we populate galaxies in halos by following \citep[also see][]{Kobayashi:2019jrn}:
\begin{itemize}
\item[(i)] {\it Central galaxies} -- a central galaxy is populated at the center of each halo with $M\ge 10^{12}~h^{-1}M_\odot$. We do not consider any off-centering between central galaxies and halo centers in this work for simplicity. We also set the velocity of the central galaxy to be equal to the velocity of the host halo. 
\item[(ii)] {\it Satellite galaxies} --  For each halo with $M\ge 10^{12}~h^{-1}M_\odot$, the number of satellite galaxies $N_{\rm sat}$ is determined from a Poisson random draw with mean given by Eq.~(\ref{eq:HOD_sat}). Once $N_{\rm sat}$ is set, we distribute each satellite galaxy according to a Navarro-Frenk-White \citep[][hereafter NFW]{nfw97} profile specified by the halo mass and the \cite{2015ApJ...799..108D} mass-concentration relation. Note that we limit the extent of the NFW profile to within the $R_{\rm 200m}$ boundary. 

For velocity assignment, we assume that satellite galaxies are randomly moving inside the host halos. Therefore, the velocities of the satellite galaxies are the sum of their host halo velocity and a random virial component. 
For this random component, we draw from a Gaussian distribution with zero mean and variance given by
\begin{align}
\left < v_x^2 \right >=\left <v_y^2 \right >=\left <v_z^2\right >=\frac{1}{3}\frac{GM_{\rm 200m}}{R_{\rm 200m}}.
\label{eq:v_sat}
\end{align}
\end{itemize}

Following \cite{2018arXiv181009456C}, we use parameter values of $M_{\rm cut}=10^{11.7}~h^{-1}M_\odot$, $M_1=10^{12.9}~h^{-1}M_\odot$, $\sigma_{\log M}=0.1$, $\kappa=1.0$, and $\alpha=1.0$ for the red-sequence galaxy mock catalog. 
This parameter configuration implies $N_{\rm cen}(M)= 1 \mbox{ for } M\ge 10^{12}h^{-1}M_\odot$, i.e. all identified halos in our halo catalogs receive a central galaxy.
The resulting galaxy number density of our mock catalogs is about $7.4\times 10^{-3}~(h^{-1}{\rm Mpc})^{-3}$ on average, which is roughly consistent with the number density of red galaxies used to identify the SDSS redMaPPer clusters. 

In addition, we generate a galaxy mock catalog for SDSS LOWZ galaxies with $M_{\rm cut}=10^{13.25}~h^{-1}M_\odot$, $M_1=10^{14.18}~h^{-1}M_\odot$, $\sigma_{\log M}=0.70$, $\kappa=1.04$, and $\alpha=0.94$ following \cite{parejko_etal2012}.

We perform this procedure on 19 independent realizations of the $(1~{h^{-1}{\rm Gpc}})^3$ box simulations.

\subsection{Cluster Finder and Mock Cluster Catalogs}
\label{sec:sim:mocks_cluster}

With the mock red-sequence galaxies in hand, we construct the cluster mock catalog using the cluster finder based on the redMaPPer cluster finder \citep{Rykoff_etal2014,Rozo2014, Rozo2015, Rozo2015_2}. The details of the algorithm and its implementation are described in the following  \cite{SunayamaMore} and \cite{Sunayama_etal2020}.

At first, we consider all the galaxies in the catalog as potential cluster central galaxies with a probability $p_{\rm free}=1$.
$p_{\rm free}$ is the prior that the galaxy does not belong to any other richer galaxy cluster, and a membership probability $p_{\rm mem}$ is the probability to be a member of the cluster.
$p_{\rm mem}$ and $p_{\rm free}$ have a simple relation of $p_{\rm free}=1-p_{\rm mem}$.
We model the photometric redshift uncertainty by assuming the specific projection length $d_{\rm proj}$. 
This is the simplifying assumption that the photometric redshift filter used to group galaxies along the redshift direction will have a poor resolution and therefore identify galaxies within a certain distance along the LOS as cluster members.
Then, we compute the initial richness $\lambda$ for each candidate central galaxy by taking all the galaxies within a radius of $0.5\mpch$ and  the LOS length $|\pi| < d_{\rm proj}$. 
We use $d_{\rm proj}=120h^{-1}{\rm Mpc}$ as our default choice, but we generate cluster mock catalogs with $d_{\rm proj}=30h^{-1}{\rm Mpc}$ and $d_{\rm proj}=60h^{-1}{\rm Mpc}$ as well.
In this first step, the membership probability will not be changed even after the galaxy is assigned to one of the cluster candidates and the cluster finder continues to go down the list of galaxy centers to find overall over-density regions.
Once the first iteration is done, we eliminate all the cluster candidates with $\lambda <3$ from the list of the candidates.
Then, we rank-order the clusters in a descending order based on the initial richness $\lambda$ and take percolation steps iteratively.  
Starting from the cluster with the largest richness, we take the following steps.
\begin{enumerate}
\item Given the $i^{\rm th}$ central galaxy in the list, recompute $\lambda$ and the membership probability based on the percolated galaxy catalog.
\item Compute the radius of $R_{c}(\lambda)$ and take all the galaxies within the radius of $R_{c}(\lambda)$ and the projection length $d_{\rm proj}$. The radial cut scales with $\lambda$ is defined as
\begin{equation}
R_{\rm c}(\lambda) = R_0 (\lambda/100.0)^{\beta}\,
\label{eq:radius}
\end{equation}
where $R_0=1.0\mpch$ and $\beta=0.2$ as adopted in redMaPPer. 

\item Among all the member galaxies, if there is a more massive central galaxy than the currently considered one (i.e., the central galaxy in a more massive halo), check whether that galaxy is already considered as a cluster center or not. If not, consider that central galaxy as a new cluster center and recompute $\lambda$ and $R_{c}(\lambda)$.
\item Determine the membership probability by numerically solving Eqn. \ref{eq:pmem} and \ref{eq:lambda}, using all the galaxies within the radius of $R_{c}(\lambda)$ and the projection length $d_{\rm proj}$. 
The membership probability is defined as
\begin{equation}
\pmem = \frac{\lambda u(x|\lambda)}{\lambda u(x|\lambda) + b(x)}\,
\label{eq:pmem}
\end{equation}
where $x$ denotes the projected distance of the galaxy from the cluster center and $b(x)$ denotes the background contamination, which is assumed to be a constant to model the uncorrelated galaxies in the foreground and the background. 
We use the projected NFW profile \citep{nfw97,Bartelmann96} for $u(x|\lambda)$ and the profile is truncated smoothly at a projected radius $R=R_{\rm c}$ with an error function as described in \cite{Rykoff_etal2014}. 
The richness $\lambda$ is defined as
\begin{equation}
\lambda = \sum_{R<R_{\rm c}(\lambda)} p_{\rm free} \pmem(x|\lambda)\,,
\label{eq:lambda}
\end{equation}
where the sum goes over all members of a galaxy cluster within a cluster radius $R<R_{\rm c}(\lambda)$ and the LOS separation $|\pi|<d_{\rm proj}$. 

\item Update the probability $p_{\rm free}$ for each galaxy to be $p_{\rm free}(1-\pmem)$ based on their membership probabilities of the current cluster. If $p_{\rm free}>0.5$, then these galaxies are eliminated from the list.
\item Repeat the steps for the next galaxy cluster in the ranked list.
\end{enumerate}

By running this cluster finder to the mock red galaxy catalog, we generate the mock cluster catalogs.
We select clusters with richness $20\leq \lambda \leq 200$ at $z=0.25$.

\begin{figure*}
    \includegraphics[width=0.43\textwidth]{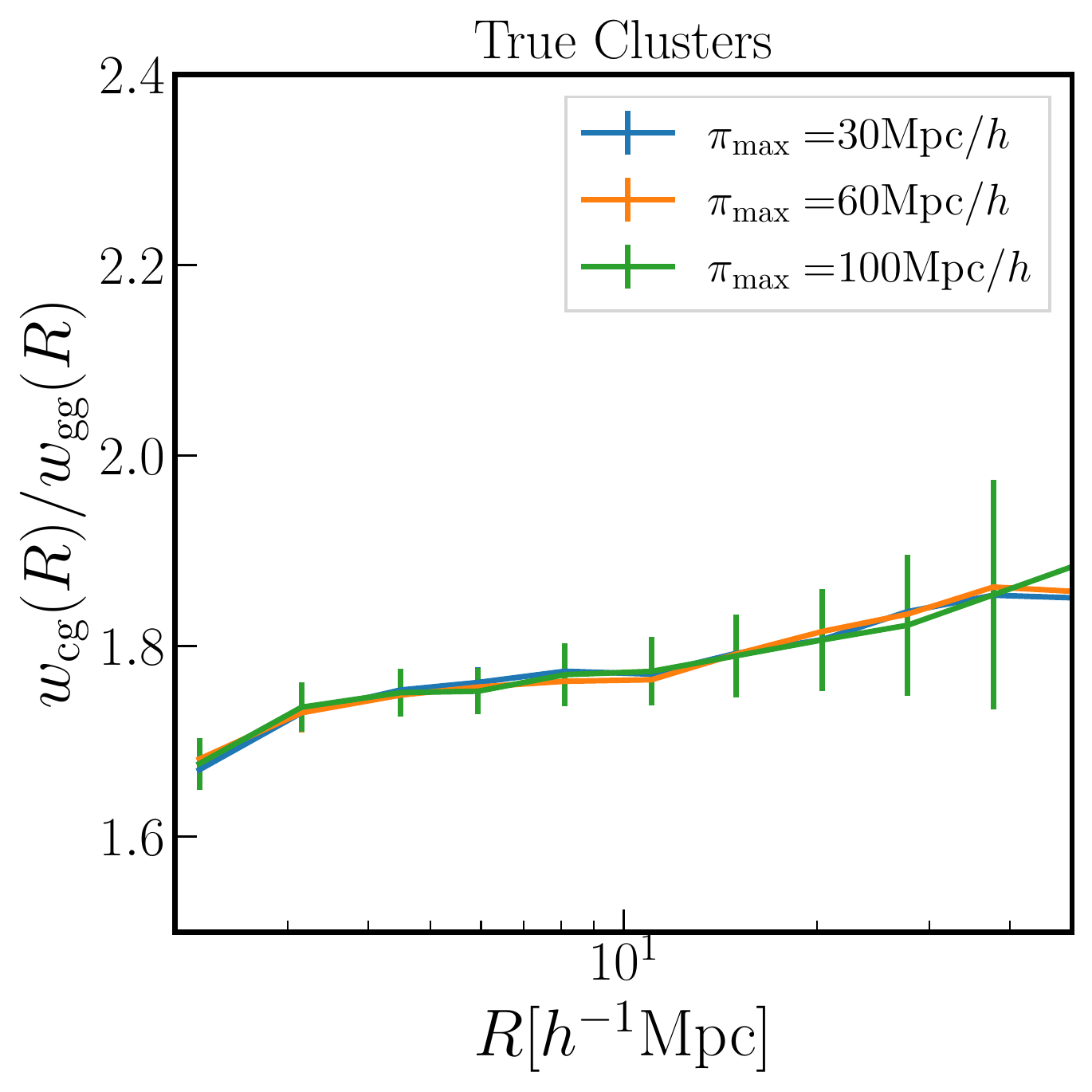}
    \includegraphics[width=0.43\textwidth]{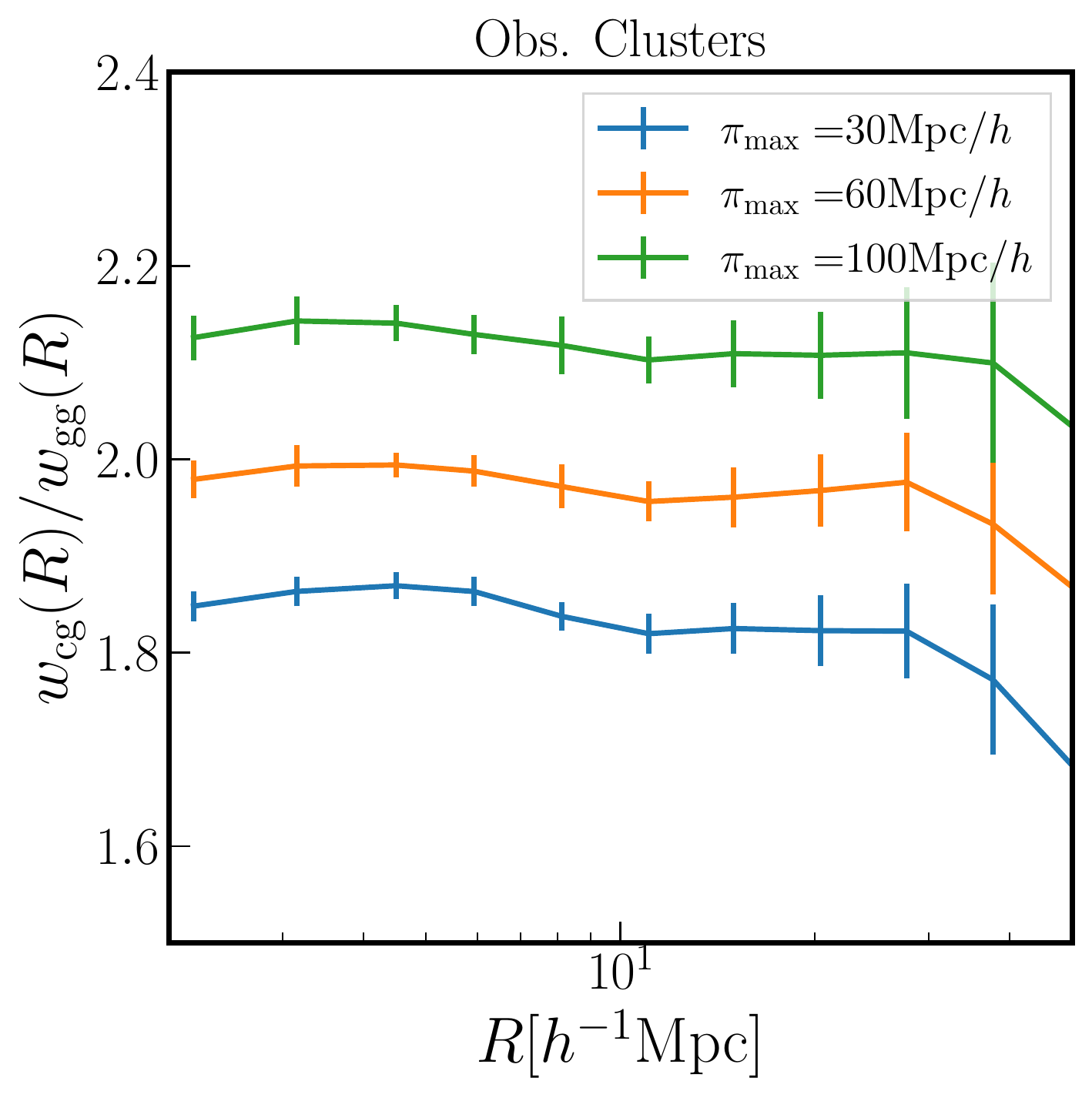}
    \caption{\label{fig:example}
    {\it Left}: 
    The projected cross-correlation function of clusters and galaxies relative to the galaxy auto-correlation function with $\pi_{\rm max}=30$, $60$, and $100h^{-1}{\rm Mpc}$. We use the halos whose masses are greater than $10^{14}h^{-1}{\rm M}_{\odot}$ (i.e., no projection effects).
    {\it Right}:
    The same as the left figure, the cross-correlation functions with the "observed" clusters which are identified by the cluster finder explained in Sec.~\ref{sec:sim:mocks_cluster}. When the clusters suffer from the projection effects, the clustering amplitude depends on the choice of the integral scale $\pi_{\rm max}$.
    }
\end{figure*}

\subsection{SDSS redMaPPer galaxy clusters}
\label{sec:obs:redmapper}

We use the publicly available catalog of galaxy
clusters identified from the SDSS DR8 photometric galaxy
catalog v5.10 by the redMaPPer cluster finding algorithm \citep{Rykoff_etal2014,Rozo2014}. 
The cluster finder uses the $ugriz$ magnitudes and their
errors, to identify overdensities of red-sequence galaxies with
similar colors as galaxy clusters. 
For each cluster, the catalog contains an optical richness 
estimate $\lambda$, a entering probability $p_{\rm cen}$, 
position as well as a photometric redshift 
$z_\lambda$ and a spectroscopic redshift $z_{\rm spec}$ if 
available.
This redMaPPer cluster catalog is volume-limited up to $z=0.33$, 
and we select galaxy clusters with $20 \leq \lambda \leq 200$ at $0.1 \leq z \leq 0.33$.
We limit our cluster sample to the ones with a spectroscopic redshift. 
This is to reduce uncertainties due to the photometric redshift in the clustering measurement. In total, we have 8648 clusters, which is $81.5\%$ of all the clusters selected in the same condition with a photometric redshift.
Throughout this paper, we use the position of the most probable central galaxy in each cluster region as a proxy of the cluster center.

We also use the random catalogs provided along with the redMaPPer cluster catalog. These catalogs contain corresponding position information, redshift, richness, and a weight for each random cluster.

\subsection{BOSS DR12 LOWZ sample}
\label{sec:obs:dr12_lowz}

We will carry out a cross-correlation of the redMaPPer galaxy clusters with spectroscopic galaxies to study the effect of the projection 
effects on the redMaPPer clusters. 
We use the spectroscopic galaxies in the large-scale structure catalogs constructed from SDSS DR12 \citep{Alam:2015}. 
In particular, we will use the LOWZ sample, since it has a large overlap in redshift range as our galaxy cluster sample. We restrict ourselves to LOWZ galaxies with redshifts between $0.1 \leq z \leq 0.33$, the same redshift range as our galaxy clusters.  
The galaxy catalogs also come with associated random galaxy catalogs that we use to perform our cross-correlation analysis.

\section{Methods}
\label{sec:method}

\begin{figure*}
    \includegraphics[width=0.3\textwidth]{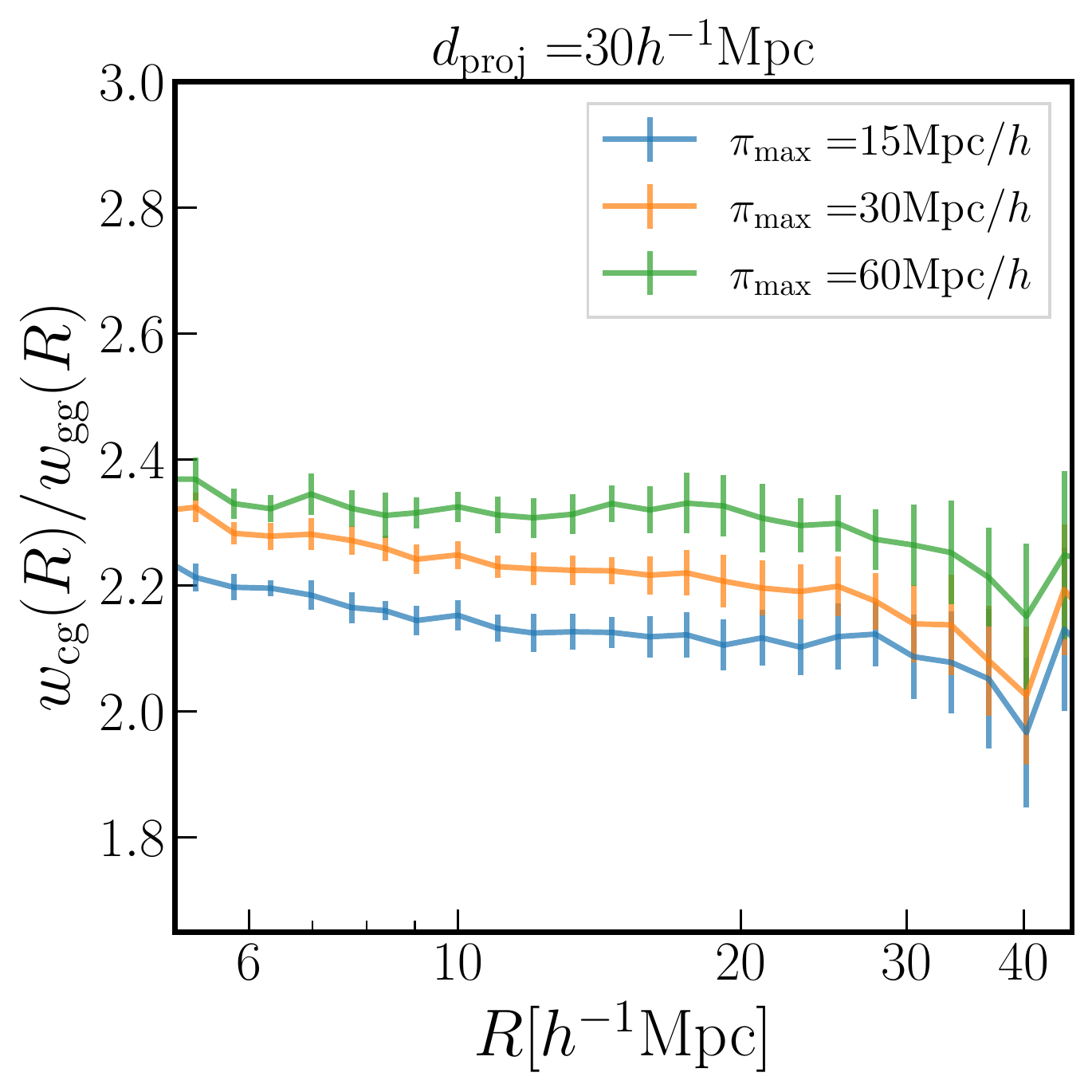}
    \includegraphics[width=0.3\textwidth]{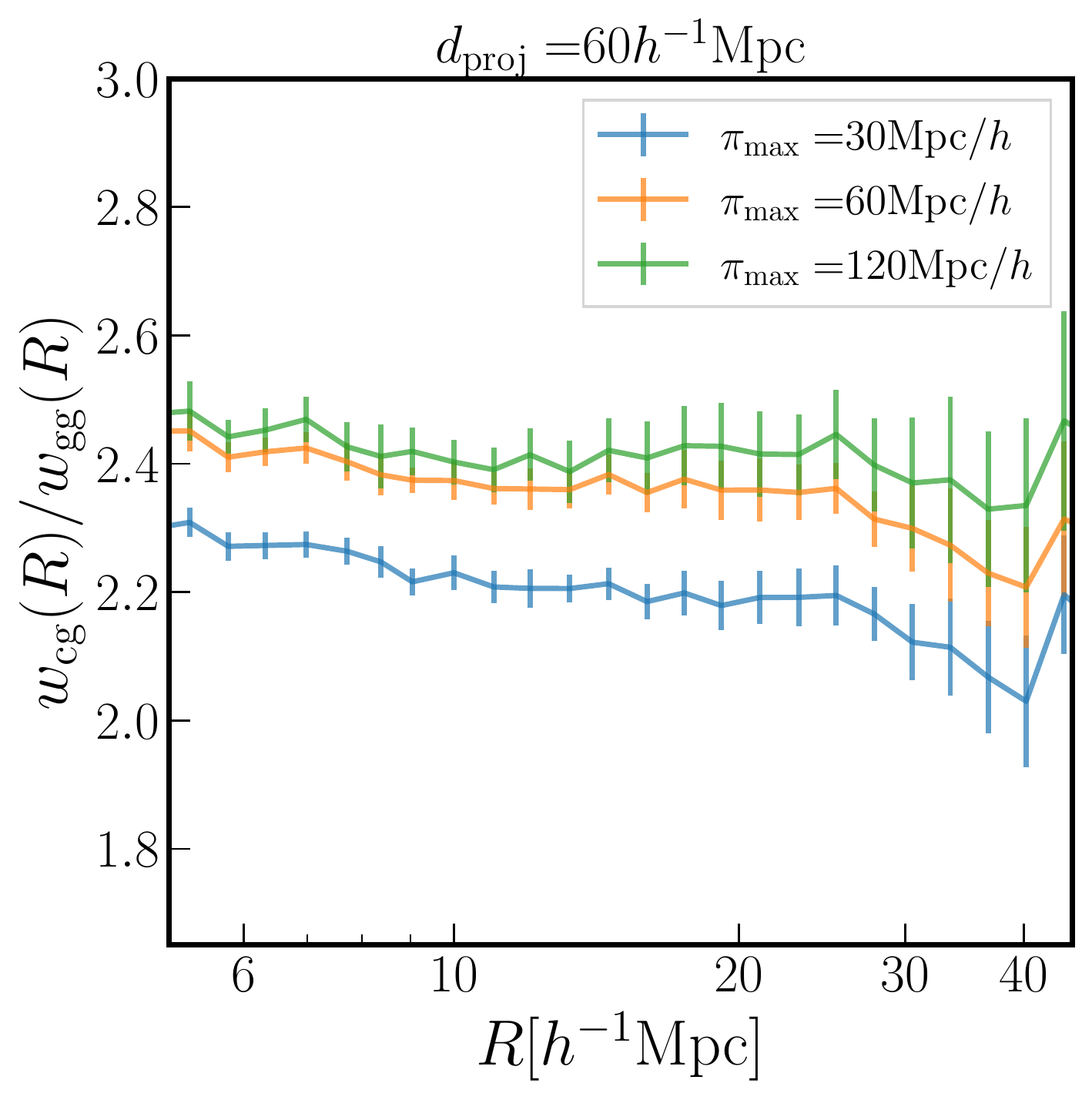}
    \includegraphics[width=0.3\textwidth]{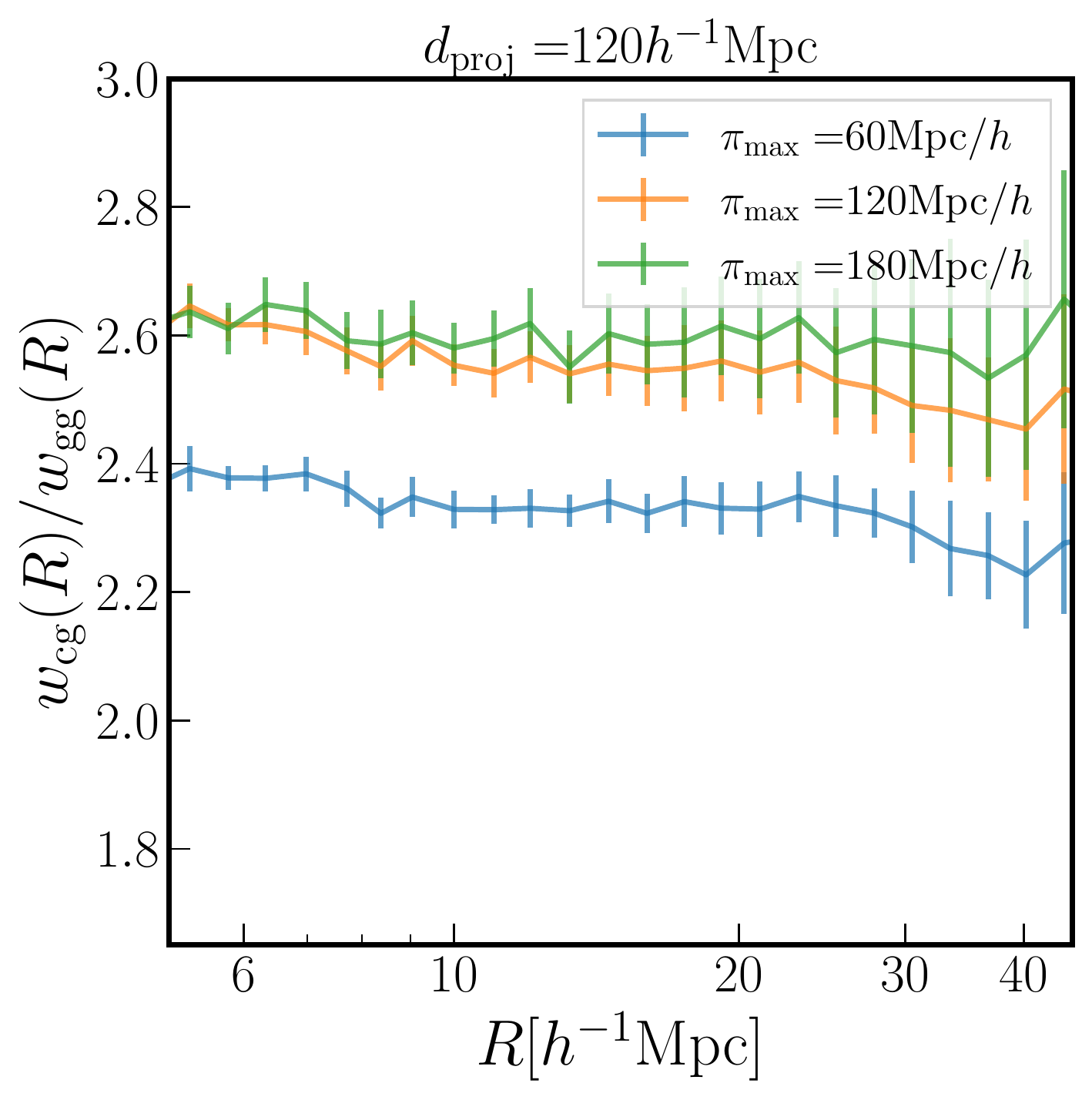}

    \includegraphics[width=0.3\textwidth]{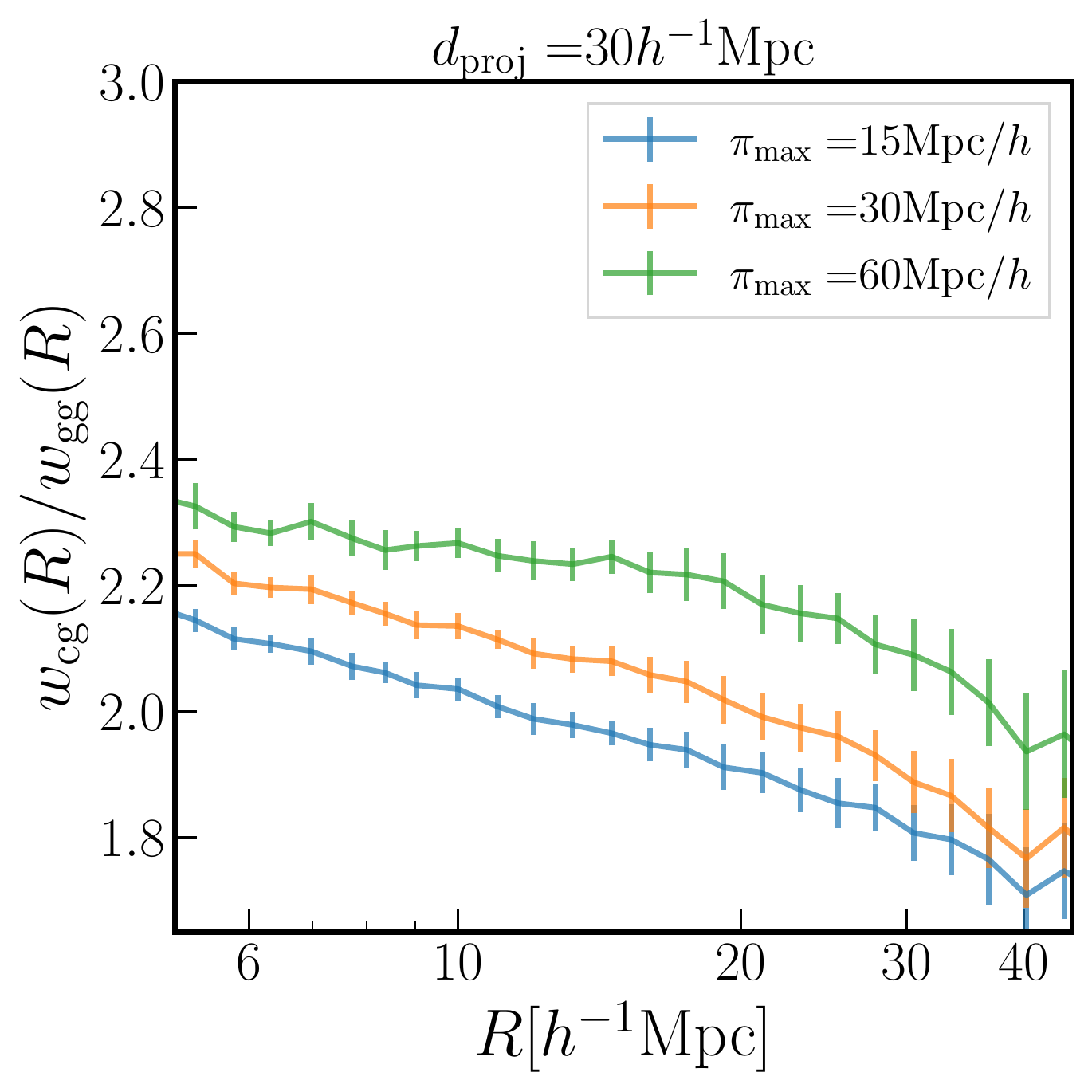}
    \includegraphics[width=0.3\textwidth]{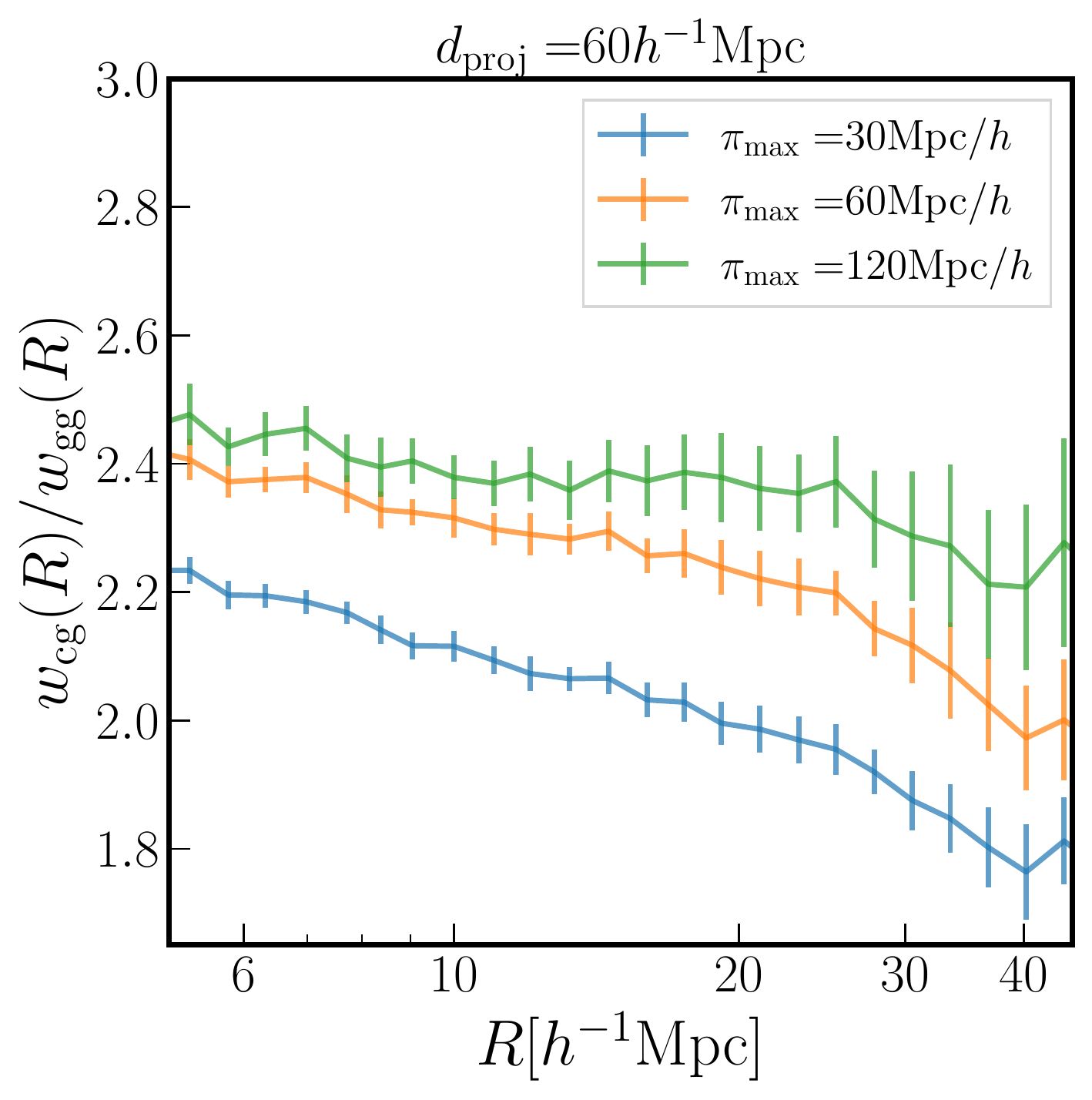}
    \includegraphics[width=0.3\textwidth]{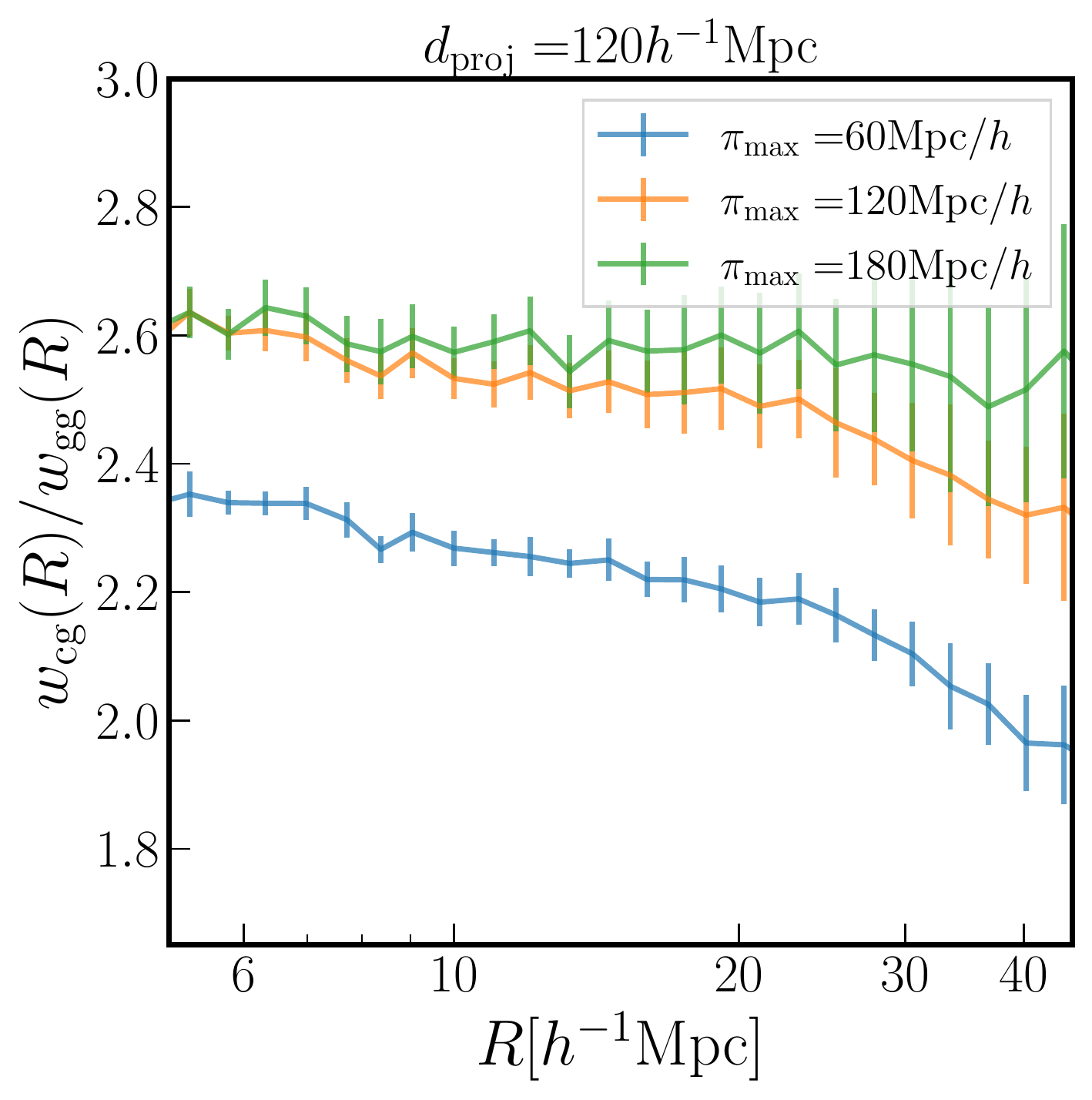}
    \caption{\label{fig:var_dproj}
    The projected cross-correlation functions of clusters and galaxies relative to the galaxy auto-correlation functions. The clusters are identified by the cluster finder with the projection length of $d_{\rm proj}=30h^{-1}{\rm Mpc}$ (left), $60h^{-1}{\rm Mpc}$ (middle), and $120h^{-1}{\rm Mpc}$ (right). The top panels are the ratios of the projected correlation functions in real-space, while the bottom panels are the ones in redshift-space. These figures indicate that the increase of clustering amplitude based on the integral scale stops when the integral scale exceeds the projection length.
    }
\end{figure*}

In this section, we introduce necessary tools to quantitatively constrain the size of the anisotropic boost in the amplitude of clustering and lensing signals caused by the projection effects.


\subsection{Projected Correlation Functions}
\label{sec:method:wp}

To measure the boost of the cluster sample, we first calculate the cluster-galaxy cross correlation function $\xi_{\rm cg}$ via the Landy-Szalay estimator \citep{LandySzalay:93}, and then compute the projected correlation function by integrating $\xi_{\rm cg}(r)$ over the LOS direction (denoted as $\pi$),
\begin{equation}
w_{\rm p,cg}(R,\pi_{\rm max})=2\int_{0}^{\pi_{\rm max}}dr_{\pi}\xi_{\rm gc}(R,r_{\pi}),\label{eq:wp}
\end{equation}
where $\pi_{\rm max}$ is the maximum integral scale.
For mocks, we simply split each $(1h^{-1}{\rm Gpc})^3$ box into eight $(0.5h^{-1}{\rm Gpc})^3$ sub-boxes and compute the covariance matrix using these sub-boxes. In total, we use 136 sub-boxes to compute the covariance matrix.
For observational data, we use 83 jackknife regions, which is about $10\times10$ square degrees corresponding to roughly $100\times100(\mpch)^{2}$ for our cluster and galaxy samples. 

Due to the anisotropic distribution of matter/galaxy around clusters, the clustering amplitude of $w_{\rm p,cg}(R)$ depends on the choice of $\pi_{\rm max}$.
As a demonstrative purpose, we use two mock cluster catalogs based on the N-body simulations described in Sec.~\ref{sec:sim:nbody} and compute the cluster-galaxy projected correlation functions. We refer the halos identified by the halo finder \texttt{Rockstar} and $M_{\rm 200m}>10^{14}\msun$ as ``True'' cluster catalog.
These cluster-sized halos do not suffer from the projection effects, because they are identified based on six-dimensional phase-space information of dark matter particles and there is no preferential selection of filamentary structure aligned to the LOS direction. 
So, the distribution of cluster-sized halos identified by \texttt{Rockstar} is isotropic.
The left panel of Fig.~\ref{fig:example} shows the ratio of $w_{\rm p,cg}/w_{\rm p, gg}$ for the case of ``True'' clusters. 
The distribution of the clusters is isotropic, and therefore the ratio does not depend on $\pi_{\rm max}$.
The right panel of Fig.~\ref{fig:example} shows the same ratio but for the clusters identified by the cluster finder.
We call the clusters identified by our cluster finder described in Sec.~\ref{sec:sim:mocks_cluster} as ``Observed'' clusters.
Due to the anisotropic distribution of matter around ``Observed'' clusters, the ratio increases as the integral scale $\pi_{\rm max}$ increases.
The reason we use the galaxy auto-correlation functions $w_{\rm p, gg}$ to extract the anisotropic boost from $w_{\rm p,cg}$ is that we need some reference objects whose distribution is isotropic.


\begin{figure*}
    \includegraphics[width=0.8\textwidth]{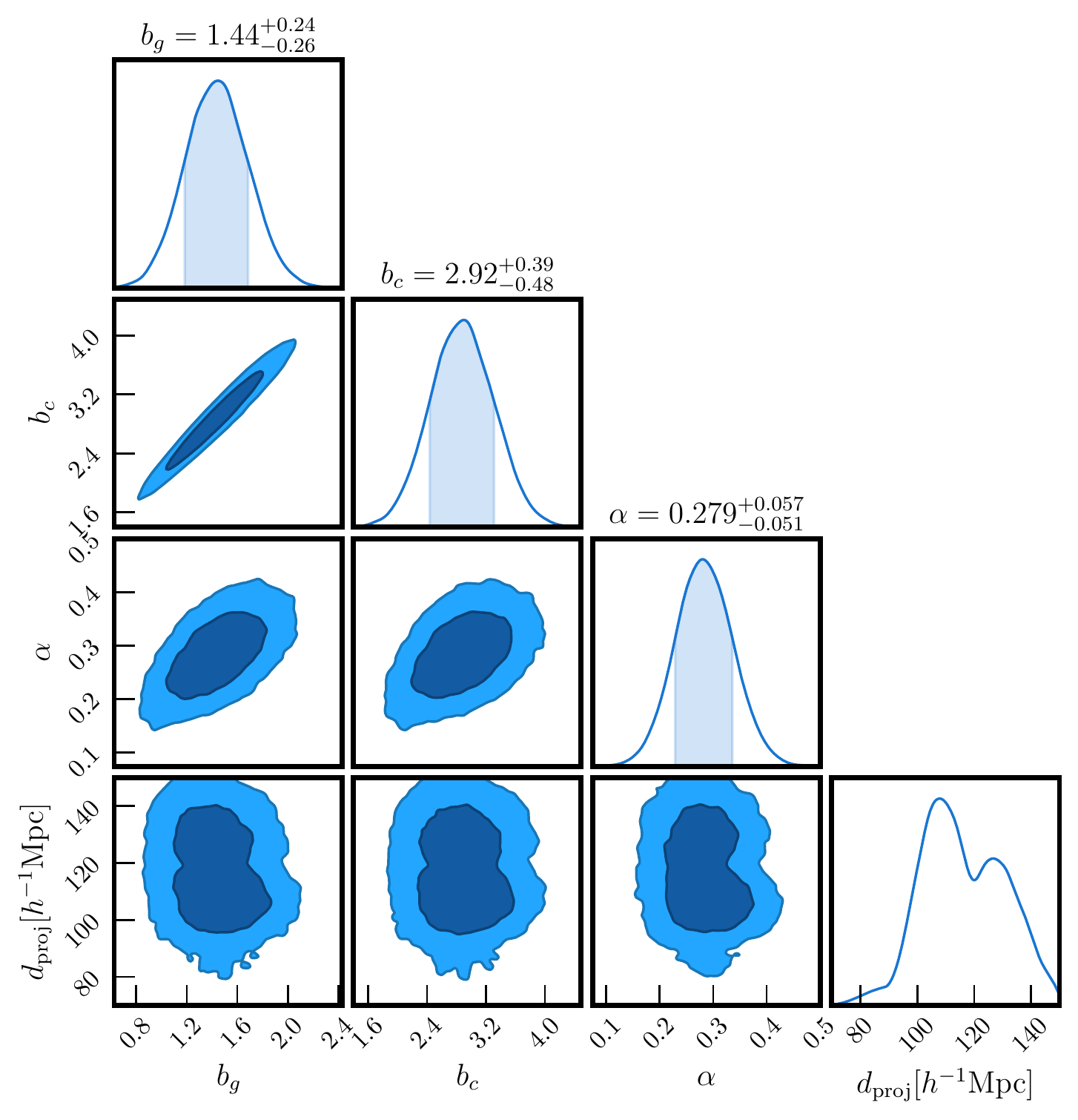}
    \caption{\label{fig:sim_mcmc_20to30}
    [The richness is from 20 to 30.]
    The posterior distribution of our model parameters using the ratio of the cross and autocorrelation functions with various integral scales. The measured parameters for the anisotropic boost parameter $\alpha$ and the projection length of $d_{\rm proj}$ are $0.283$ and $108h^{-1}{\rm Mpc}$, while the expected parameter values (shown as dotted lines) are boost=$0.273$ and $d_{\rm proj}=120h^{-1}{\rm Mpc}$.
    }
\end{figure*}

\subsection{Theoretical Modeling of Projection Effects}
\label{sec:method:model}
We model the anisotropic boost on $w_{\rm p,cg}(R)$ with two parameters $\alpha_0$ and $d_{\rm proj}$,
\begin{align}
w_{\rm p,cg}(R,\pi_{\rm max})=(1+\alpha_0(\pi_{\rm max}, d_{\rm proj}))w_{\rm iso,cg}(R)
\label{eq:inertia}
\end{align}
where $\alpha_0$ is the boost of the clustering amplitude and $d_{\rm proj}$ is the projection length of the clusters (i.e., the maximum LOS distance of the member galaxies from the cluster center).
$w_{\rm iso,cg}(R)$ is the expected projected cross-correlation function for the case of isotropically distributed clusters with the same halo masses.
We assume that the boost parameter $\alpha_0$ is scale-independent on large scales.
To model $\alpha_0$, we make the following assumptions:
\begin{enumerate}
\item $\alpha_0$ increases constantly as $\pi_{\rm max}$ increases,
\item the increase of $\alpha_0$ stops when $\pi_{\rm max}>d_{\rm proj}$.
\end{enumerate}
These two assumptions are made based on the findings in \cite{BuschWhite:17} and \cite{SunayamaMore}. Both studies investigated the cause of large assembly signals detected in \cite{Miyatake:2016}. These studies found that the projection effects can boost the clustering amplitude and therefore can give a false detection of the assembly bias.
These studies built a cluster finder with the projection effects and found that the anisotropic matter distribution around the clusters due to the projection effects can only extend up to $\pi \leq d_{\rm proj}$ (see Fig. 12 in \cite{BuschWhite:17} and Fig. 4 in \cite{SunayamaMore}).
The above two assumptions are based on these findings.

With these assumptions, we can describe $\alpha_0$ as
\begin{equation}
    \alpha_0=
    \begin{cases}
      \alpha, & \text{if}\ \pi_{\rm max}>d_{\rm proj} \\
      \alpha \frac{\pi_{\rm max}}{d_{\rm proj}}. & \text{otherwise}
    \end{cases}
  \end{equation}
The top panels of Fig.~\ref{fig:var_dproj} show the ratio of $w_{\rm p,cg}/w_{\rm p,gg}$ for "Observed" clusters with richness $20 \leq \lambda \leq 30$. Each panel shows the clusters identified with $d_{\rm proj}=30 h^{-1}{\rm Mpc}$ (left), $60 h^{-1}{\rm Mpc}$ (middle), and $120 h^{-1}{\rm Mpc}$ (right).
We compute the projected correlation functions in real-space.
This is the supportive evidence for the second assumption that the increase of the boost stops roughly after $\pi_{\rm max}>d_{\rm proj}$ for the case of $\pi_{\rm max}=60h^{-1}{\rm Mpc}$ and $120h^{-1}{\rm Mpc}$, but not for the case of $\pi_{\rm max}=30h^{-1}{\rm Mpc}$.
This means that our model is only valid for clusters with the large projection length. 
Since velocity dispersion of galaxies in clusters is roughly $\sigma_{v}\sim 3000{\rm km/s}$, we can only achieve $d_{\rm proj}=30h^{-1}{\rm Mpc}$ even for the case of spectroscopically identified clusters, and therefore assuming larger projection lengths than $30h^{-1}{\rm Mpc}$ is a fine assumption.
In real-space, the ratio of the projected correlation functions is assumed to be constant, and therefore its model requires only three parameters $b_c/b_g$, $\alpha$, and $d_{\rm proj}$: 
\begin{equation}
\frac{w_{\rm cg}(R,\pi_{\rm max})}{w_{\rm gg}(R,\pi_{\rm max})} = \frac{b_c}{b_g} (1+\alpha_0(\pi_{\rm max},d_{\rm proj})).
\label{eq:ratio_real}
\end{equation}
The projected correlation function, however, is sensitive to the redshift-space distortion (RSD) effect when a sufficiently small value of $\pi_{\rm max}$ is taken (see Fig. 6 in \cite{vdBosch2013}).
Even though we take the ratio of the two projected correlation functions, the scale-dependence induced by the RSD effect is different for the cluster-galaxy cross-correlation functions and galaxy auto-correlation functions.
The model with the RSD effect has the following additional term to Eqn.~\ref{eq:ratio_real}.
\begin{equation}
\frac{(1+\frac{\beta_c+\beta_g}{3}+\frac{\beta_c \beta_g}{5}) w_{\rm p,0}+(\frac{2(\beta_c+\beta_g)}{3}+\frac{4\beta_c \beta_g}{7})w_{\rm p,2}+\frac{8\beta_c \beta_g}{35} w_{\rm p,4}}{(1+\frac{2 \beta_g}{3}+\frac{\beta_g^2}{5})w_{\rm p,0}+(\frac{4\beta_g}{3}+\frac{4 \beta_g^2}{7})w_{\rm p,2}+\frac{8 \beta_g^2}{35} w_{\rm p,4}},
\label{eq:ratio_RSD}
\end{equation}
where $\beta_c$ and $\beta_g$ are the growth rate $f$ divided by cluster and galaxy biases $b_c$ and $b_g$ respectively, $w_{\rm p,0}(R)$, $w_{\rm p,2}(R)$, and $w_{\rm p,4}(R)$ are the monopole, quadrupole, and hexadecapole of the projected correlation function:
\begin{equation}
w_{\rm{p},n}(R) = 2\int_{R}^{\infty}\xi_n(r) L_n(\mu) \frac{r dr}{\sqrt{(r^2-R^2)}},
\label{eq:abel}
\end{equation}
where $L_n(\mu)$ is the $n$th Legendre polynomials with $\mu=\sqrt{(r^2-R^2)}/r$. Following Eqn. 52-54 in \cite{vdBosch2013}, we define $\xi_n(r)$ as
\begin{align}
\xi_0(r) &= \xi_{\rm NL}(r),\\
\xi_2(r) &= \xi_{\rm NL}(r)-3J_3(r),\\
\xi_4(r) &= \xi_{\rm NL}(r)+\frac{15}{2}J_3(r)-\frac{35}{2}J_5(r),
\label{eq:xi_n}
\end{align}
where $\xi_{\rm NL}(r)$ is the non-linear matter correlation function and 
\begin{align}
J_n(r) =\frac{1}{r^n}\int^{r}_0 \xi_{\rm lin}(y)y^{n-1}{\rm d}y.
\label{eq:J_n}
\end{align}

The bottom panels of Fig.~\ref{fig:var_dproj} show the ratio $w_{\rm cg}(R)/w_{\rm gg}(R)$ in redshift-space for the case of $d_{\rm proj}=30h^{-1}{\rm Mpc}$, $60h^{-1}{\rm Mpc}$, and $120h^{-1}{\rm Mpc}$.
Compared to the case of real-space (top panels), the ratio $w_{\rm cg}(R)/w_{\rm gg}(R)$ in redshift-space shows a stronger scale-dependence due to the RSD effect, and the second assumption (i.e., the increase of the boost stops roughly after $\pi_{\rm max}>d_{\rm proj}$) seems violated. 
However, as long as the second assumption holds in real-space, this seeming violation of the assumption can be modeled following \cite{vdBosch2013} and will not be a problem.
For the rest of the paper, we use the mock cluster catalog with $d_{\rm proj}=120h^{-1}{\rm Mpc}$ if it is not specified.

\subsection{Parameter Inference}
\label{sec:method:mcmc}
We assume a Gaussian likelihood model and compute the likelihoods with the measurements $\mathbf{d}$ and the model predictions $\bm\upmu(\bm\uptheta)$, where $\bm\uptheta$ are the parameters, with covariances $\mathbf{C}$:
\begin{equation}
    \ln \mathcal{L}(\mathbf{d}|\bm\uptheta) = 
    -\frac{1}{2} \left[\mathbf{d}-\bm{\upmu}(\bm\uptheta)\right]^\intercal \mathbf{C}^{-1} \left[\mathbf{d}-\bm{\upmu}(\bm\uptheta)\right].
\end{equation}
The data vector $\mathbf{d}$ is given by the ratios $w_{\rm p cg}/w_{\rm p,gg}$ for $\pi_{\rm max}=30, 60, 90, 120, 150 h^{-1}{\rm Mpc}$ on scales $R$.
The parameters used for the model are $b_c$, $b_g$, $\alpha$, and $d_{\rm proj}$.
We use flat priors for $\alpha$ and $d_{\rm proj}$: $\alpha \in [0,1]$, and $d_{\rm proj} \in [30,150]$. 
For $b_g$ and $b_c$, we use Gaussian priors: $b_g \sim \mathcal{N}(1.5,1.)$ and $b_c \sim \mathcal{N}(3.,1.)$.
This is because the values of $b_g$ and $b_c$ are constrained independently only from the difference in the scale-dependence due to the RSD effect, and therefore these parameters are not strongly constrained with flat priors.

To estimate the covariance matrix, we use roughly 80 independent realizations either by sub-dividing a set of the simulation realizations or by using the jackknife resampling method. 
Due to the large size of our data vector, the number of realizations or the jackknife regions is not enough to obtain the unbiased inverse covariance matrix. 
To mitigate this issue, we use a Principal Component Analysis on the covariance matrix $\mathbf{C}$ following the way described in Appendix C7 of \cite{Behroozi2019}.
We first diagonalize the covariance matrix through an orthoganal matrix $\mathbf{U}$, $\mathbf{C}=\mathbf{U}\mathbf{D}\mathbf{U}^{-1}$ where $\mathbf{D}={\rm diag}(\sigma_1,\sigma_2,...)$, and then replace the $i$-th diagonal element $\sigma_i$ with the effective error $\sigma_{\rm eff,i}={\rm max}(0.1,\sigma_i)$ for the case of mock analysis.
This is to include the systematic errors due to observational measurements such as fiber collisions and edge effects.
The corresponding data vector for this diagonal covariance matrix $\mathbf{D}$ is $\mathbf{U}^{-1}(\mathbf{d}-\bm{\upmu}(\bm\uptheta))$.

We perform Bayesian parameter inferences with the above likelihoods to obtain the posterior distribution of the parameters. 
To do that, we use the affine invariant Monte Carlo Markov Chain (MCMC) ensemble sampler \textit{emcee} \citep{emcee}.

\section{Results}
\label{sec:results}

In this section, we first show the results from the mock cluster catalog described in Sec.~\ref{sec:sim:mocks_galaxy} and validate our model of the projection effects. Then, we apply our model to SDSS redMaPPer cluster catalog (details in Sec.~\ref{sec:obs:redmapper}). 

\subsection{Simulations}
\label{sec:res:sim}

We show the derived posterior constraints of our model parameter for the mock cluster catalog with $20 \leq \lambda \leq 30$ and $d_{\rm proj}=120^{-1}{\rm Mpc}$ in Fig.~\ref{fig:sim_mcmc_20to30}. 
For this analysis, we use the ratios $w_{\rm p cg}/w_{\rm p,gg}$ for $\pi_{\rm max}=30, 60, 90, 120,$ and $150 h^{-1}{\rm Mpc}$ on scales $12h^{-1}{\rm Mpc} \leq R \leq 40h^{-1}{\rm Mpc}$ with 13 bins each, and the covariance matrix is computed from 136 $(0.5h^{-1}{\rm Mpc})^3$ sub-boxes.
The diagonal panels show the one-dimensional marginalized posterior distributions of each of the four parameters and the contours in the off-diagonal panels are the $1 \sigma$ confidence region for each of the parameter combinations. 

Since the value of $\alpha$ is not an input parameter, we measured it using a different method to validate our model. 
Using the same cluster mock catalog, we first measured the lensing signal and compared it with the prediction from the emulator \textit{darkemu} \citep{darkemu}. \textit{darkemu} takes the halo masses of the clusters as an input and gives the predicted lensing signal. Since \textit{darkemu} assumes that the distribution of halos is isotropic, the deviation of the measured lensing signal from the prediction is the size of the anisotropic boost.
The measured value of $\alpha$ from the lensing signal is $\alpha=27.6\pm{4.7}\%$.
The details of the measurements and the results for other cluster catalogs with different richness bins and other projection lengths are in Appendix~\ref{sec:appA}.

Fig.~\ref{fig:sim_mcmc_20to30} shows that the posterior distribution of $\alpha$ from our method nicely agrees with the measured value of $\alpha$ from lensing. 
However, the posterior distribution of $d_{\rm proj}$ is not well-constrained and the best-fit value is smaller than the expected value of $d_{\rm proj}=120h^{-1}{\rm Mpc}$.
This may be because the second assumption (i.e., the increase of the amplitude ratio stops at $\pi_{\rm max}=d_{\rm proj}$) is too simple. 
However, the important thing here is not to precisely constrain the value of $d_{\rm proj}$, but rather to get some rough ideas of how long the distribution of member galaxies is extended along the LOS.
At least, the posterior distribution of $d_{\rm proj}$ is not significantly under/over-estimating $d_{\rm proj}$. Further improvements to constrain $d_{\rm proj}$ on our model will be future work.

\begin{figure}
    \includegraphics[width=0.45\textwidth]{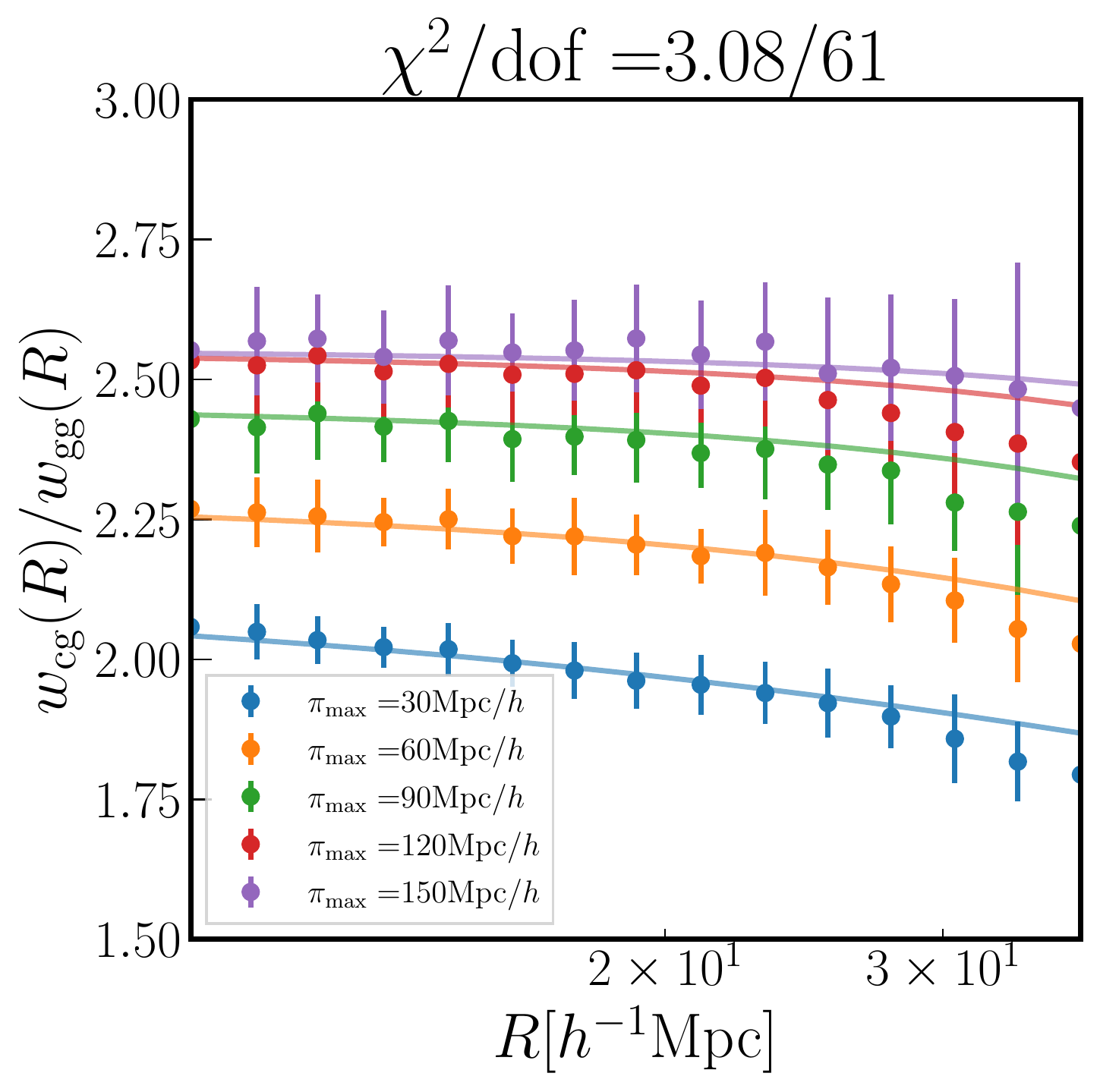}
    \caption{\label{fig:sim_pred_20to30}
    The projected cross-correlation functions of clusters and galaxies relative to the galaxy auto-correlation functions with $\pi_{\rm max}=30, 60, 90, 120$, and $150h^{-1}{\rm Mpc}$. The circles with error bars are based on the mock measurements and the solid lines are the predictions based on our model with the best-fit parameters.
    }
\end{figure}

\begin{figure}
    \includegraphics[width=0.45\textwidth]{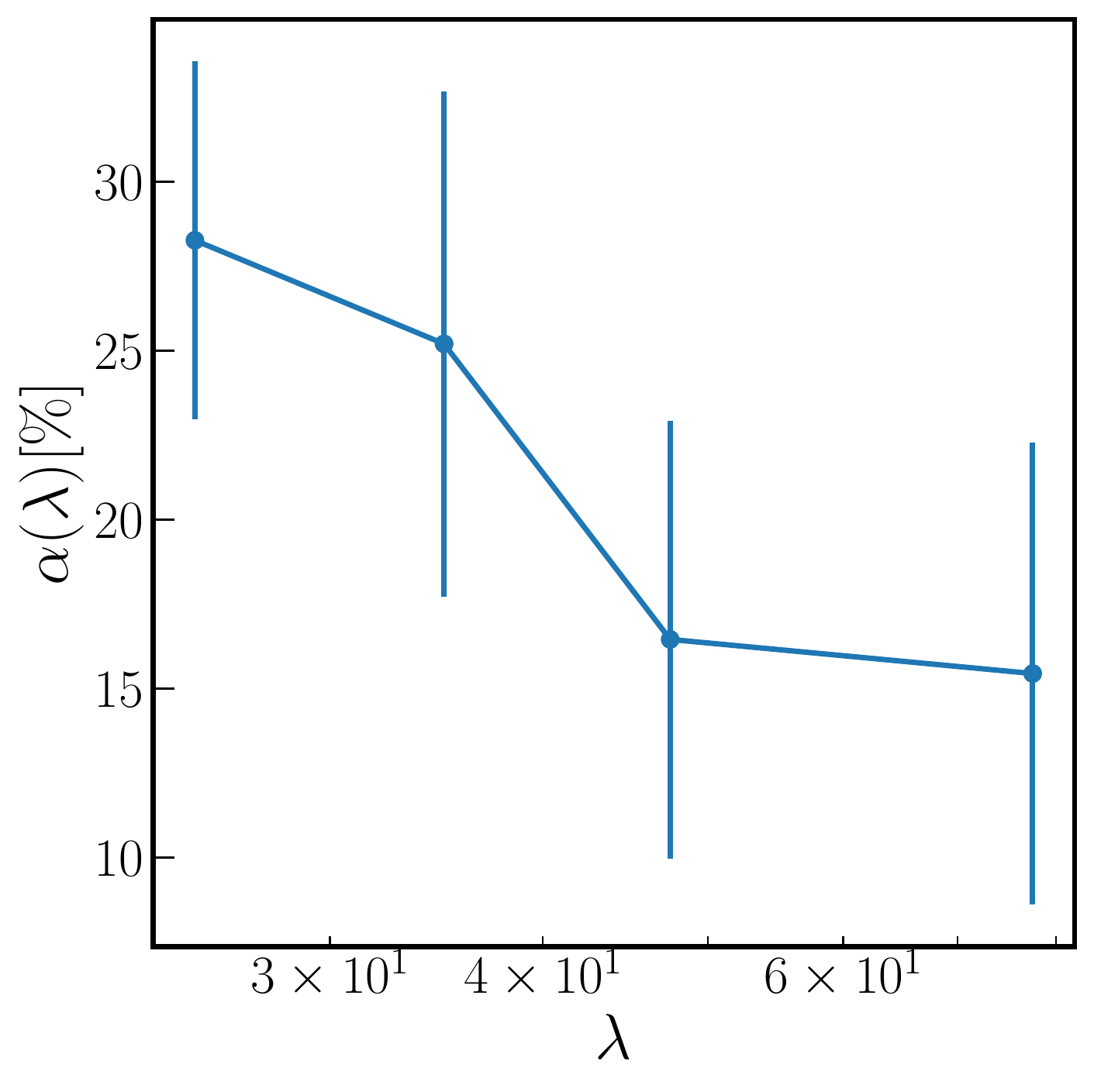}
    \caption{\label{fig:alphas_mock}
    The best-fit values of the anisotropic boost parameter $\alpha$ as a function of richness $\lambda$. As richness increases, the value of $\alpha$ decreases (i.e., smaller anisotropic boost due to the projection effects).
    }
\end{figure}

\begin{figure*}
    \includegraphics[width=0.8\textwidth]{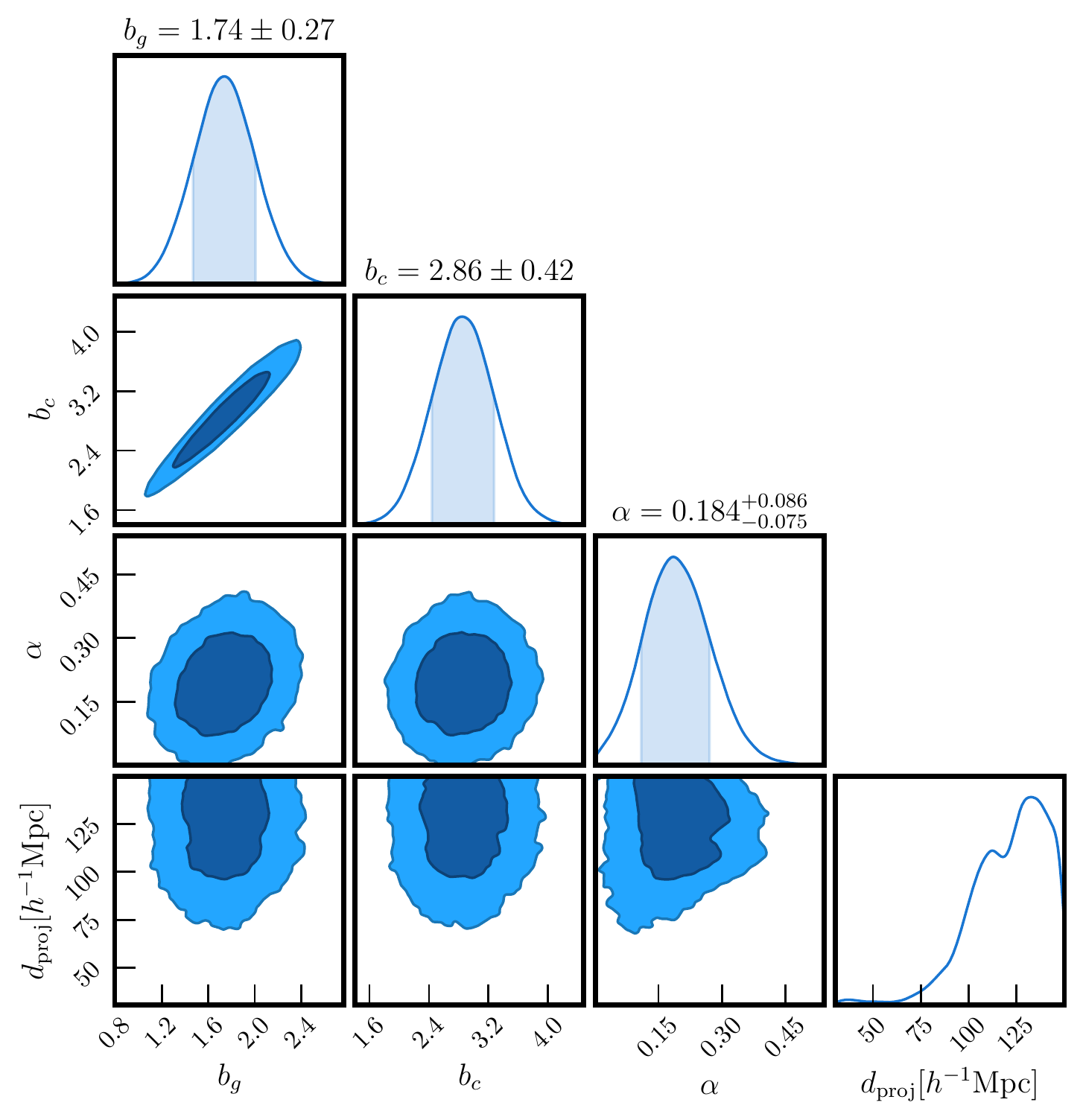}
    \caption{\label{fig:obs_mcmc_20to30}
    The same figures as Fig.~\ref{fig:sim_example} but using SDSS redMaPPer clusters and DR12 galaxies at $0.1 \leq z \leq 0.33$ with richness $20 \leq \lambda \leq 30$.
    The posterior distribution of cluster projection model parameters given the ratio of the cross and autocorrelation functions with various integral scales. The measured parameters for boost and $d_{\rm proj}$ are 0.16 and $136h^{-1}{\rm Mpc}$.
    }
\end{figure*}

Fig.~\ref{fig:sim_pred_20to30} compares the amplitude ratio of $w_{\rm cg}(R)/w_{\rm gg}(R)$ predicted by the best-fit values from the posterior distributions (solid line) to the mock measurements (circles with error bars).
Overall, the model predictions well describe the data points from the mock measurement. 
However, the scale-dependent suppression on large scales is slightly stronger for the measurements than the prediction from the RSD effects. 
This might be due to unmodeled correlations of the clusters with tidal fields.
The reason the $\chi^2$ value is so small despite the large degree of freedom (=61) is that we use the covariance matrix including observational systematic errors, which are not present in the mock data vector.

Fig.~\ref{fig:alphas_mock} shows the anisotropic boost $\alpha$ as a function of richness $\lambda$. For larger richness clusters, the value of $\alpha$ gets smaller. This is mainly because the fraction of the "projected" clusters, whose member galaxies are mostly interlopers, gets smaller for more massive clusters. \cite{Sunayama_etal2020} showed that these clusters are the cause of the anisotropic boost and the size of the boost mostly depends on the fraction of these "projected" clusters.

\subsection{Observations}
\label{sec:res:obs}
In this section, we present the results using the SDSS redMaPPer clusters and LOWZ spectroscopic galaxies. 
As is described in Sec.~\ref{sec:obs:redmapper} and \ref{sec:obs:dr12_lowz}, we select clusters and galaxies at $0.1 \leq z \leq 0.33$ and measure the cluster-galaxy projected correlation functions as well as galaxy auto-correlation functions with $\pi_{\rm max}=30, 60, 90, 120,$ and $150h^{-1}{\rm Mpc}$ on scales $10h^{-1}{\rm Mpc} \leq R \leq 42 h^{-1}{\rm Mpc}$ with 8 bins each.
To look for the features of the projection effect in the ratio of $w_{\rm p,cg}(R)/w_{\rm p,gg}(R)$, we limit to use the redMaPPer clusters with spectroscopic redshifts.
This is because the photometric redshift uncertainties dilute the dependence of the ratio on the choice of integral scale $\pi_{\rm max}$.
We use 83 jackknife regions in order to compute the error in the measurements and its covariance matrix. The typical size of each of these jackknife patches is about $10 \times 10$ square degrees, which corresponds to roughly $100 \times 100 (h^{-1}{\rm Mac})^2$ for our cluster and galaxy samples. Note that we do include the correction factors in \cite{Hartlapetal2007}.

\begin{figure}
    \includegraphics[width=0.45\textwidth]{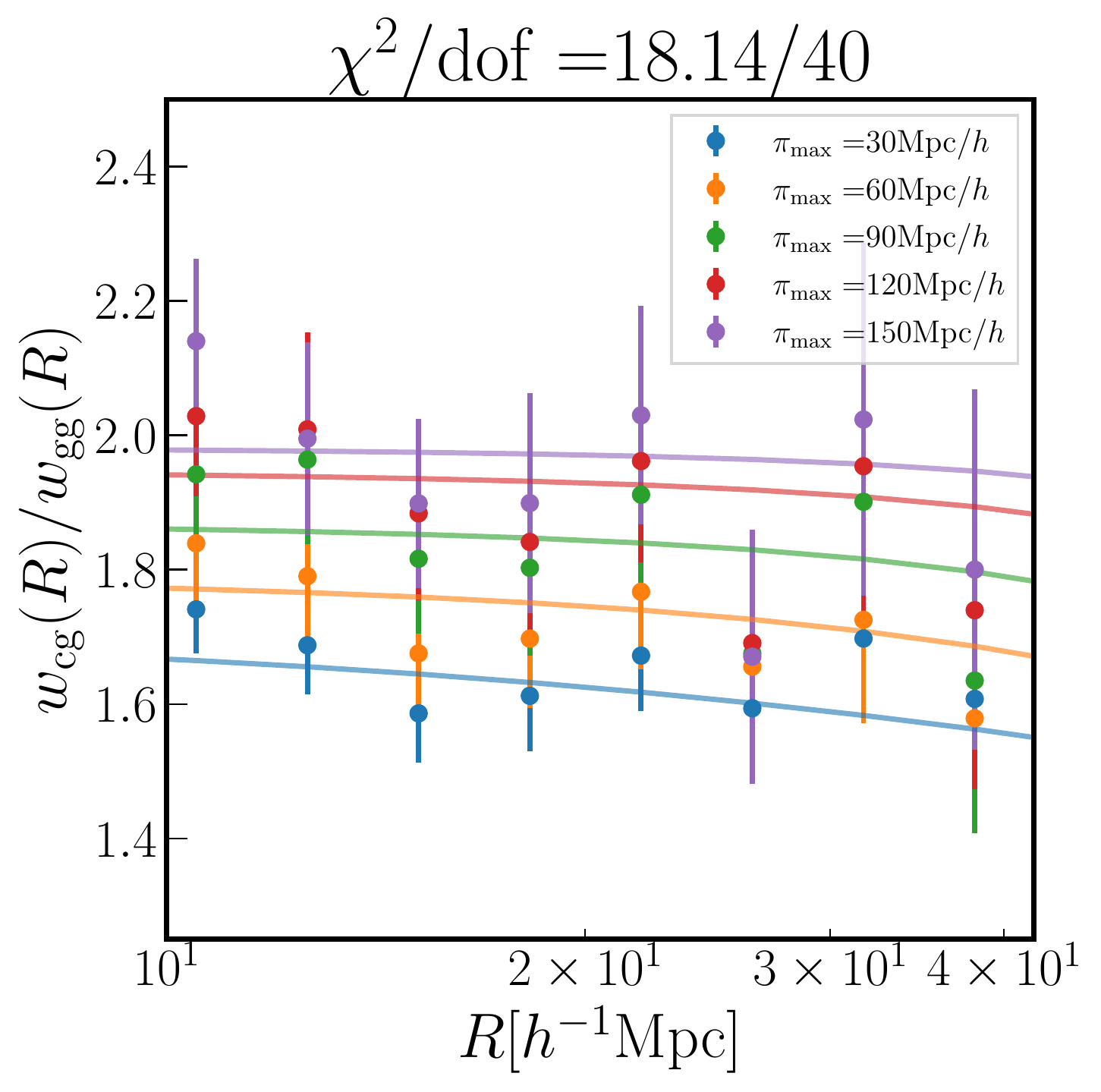}
    \caption{\label{fig:obs_pred_20to30}
    The projected cross-correlation functions of SDSS RM clusters and galaxies relative to the galaxy auto-correlation functions with $\pi_{\rm max}=30, 60, 90, 120$, and $150h^{-1}{\rm Mpc}$ with $20 \leq \lambda \leq 30$. The circles with error bars are from the measurements and the solid lines are the predictions based on our model with the best-fit parameters.
    }
\end{figure}

Fig.~\ref{fig:obs_mcmc_20to30} shows the derived posterior constraints of our model parameters for the redMaPPer clusters with richness $20 \leq \lambda \leq 30$.
The anisotropic boost factor $\alpha$ is constrained to $\alpha=18.4\pm 8.6\%$, which is consistent with the values from \cite{Park_etal2021} and \cite{To_Krause2021}.
The predicted value for $d_{\rm proj}$ is $d_{\rm proj}=136h^{-1}{\rm Mpc}$. Due to the limited range of $d_{\rm proj}$ up to $150h^{-1}{\rm Mpc}$, our posterior is truncated on the large $d_{\rm proj}$ end, but the size of the 1$\sigma$ error on the small $d_{\rm proj}$ is roughly $\Delta d_{\rm proj}=30h^{-1}{\rm Mpc}$. 
This large $d_{\rm proj}$ does not mean that the structure of all the clusters is extended to $136h^{-1}{\rm Mpc}$, but rather member galaxies of some clusters are distributed to roughly $136h^{-1}{\rm Mpc}$ along the LOS.
While the simulation analysis in Fig.~\ref{fig:sim_mcmc_20to30} shows a weak degeneracy between $\alpha$ and $b_c$/$b_g$, Fig.~\ref{fig:obs_mcmc_20to30} shows no degeneracy between these parameters.
This is because the scale-dependence of $w_{\rm cg}(R)/w_{\rm gg}(R)$ is weaker for the observational data than the mock data as shown in Fig.~\ref{fig:obs_pred_20to30}.

\begin{figure}
    \includegraphics[width=0.45\textwidth]{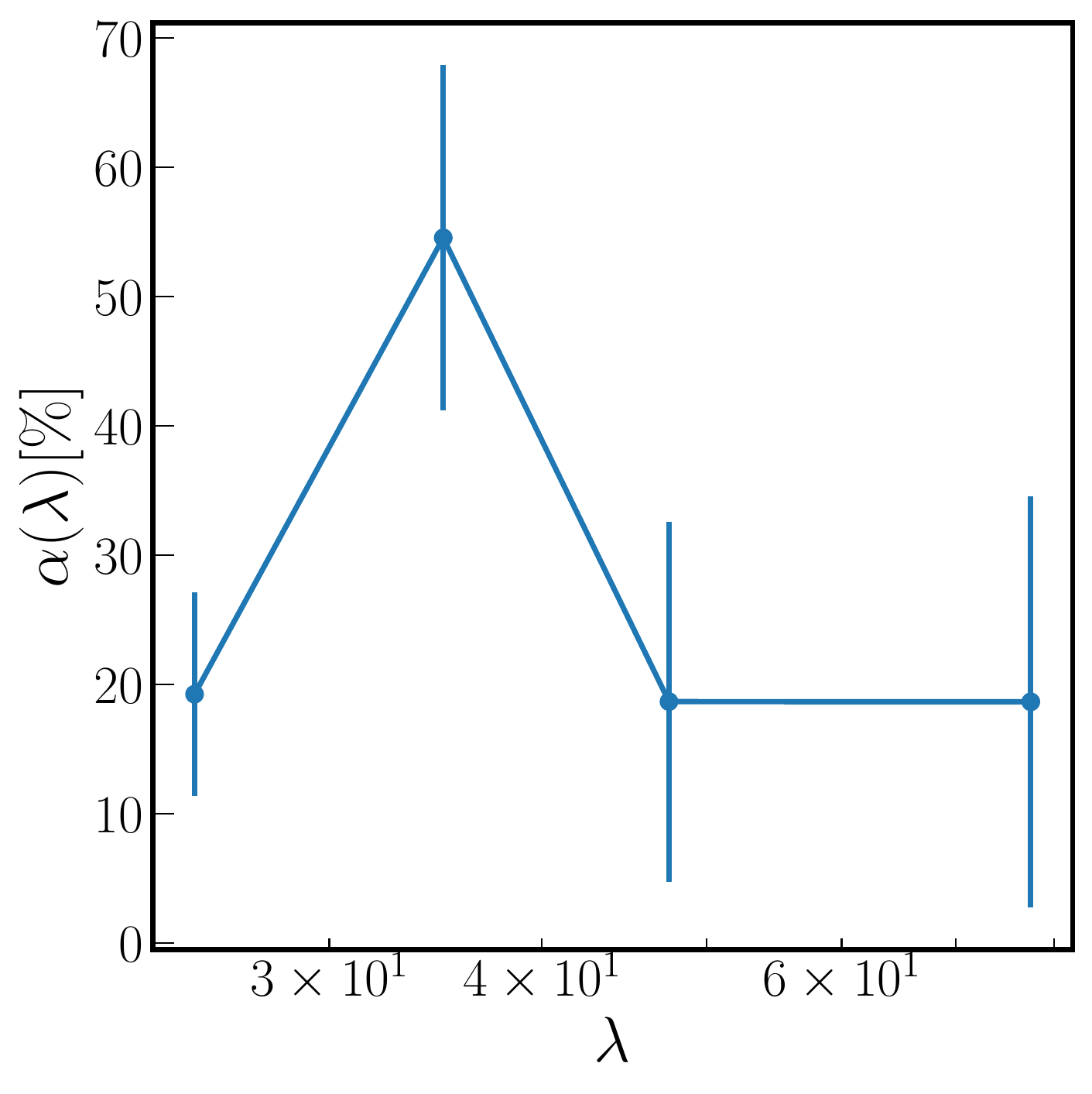}
    \caption{\label{fig:alphas_obs}
    The best-fit values of the anisotropic boost parameter $\alpha$ as a function of richness $\lambda$. As richness increases, the value of $\alpha$ decreases (i.e., smaller anisotropic boost due to the projection effects) except for the cluster sample with $30 \leq \lambda \leq 40$.
    }
\end{figure}

Fig.~\ref{fig:obs_pred_20to30} compares the amplitude ratio of $w_{\rm cg}(R)/w_{\rm gg}(R)$ predicted by the best-fit values from the posterior distributions (solid line) to the observational measurements (circles with error bars).
Overall the model predictions agree well with the measurements. 

\begin{figure}
    \includegraphics[width=0.45\textwidth]{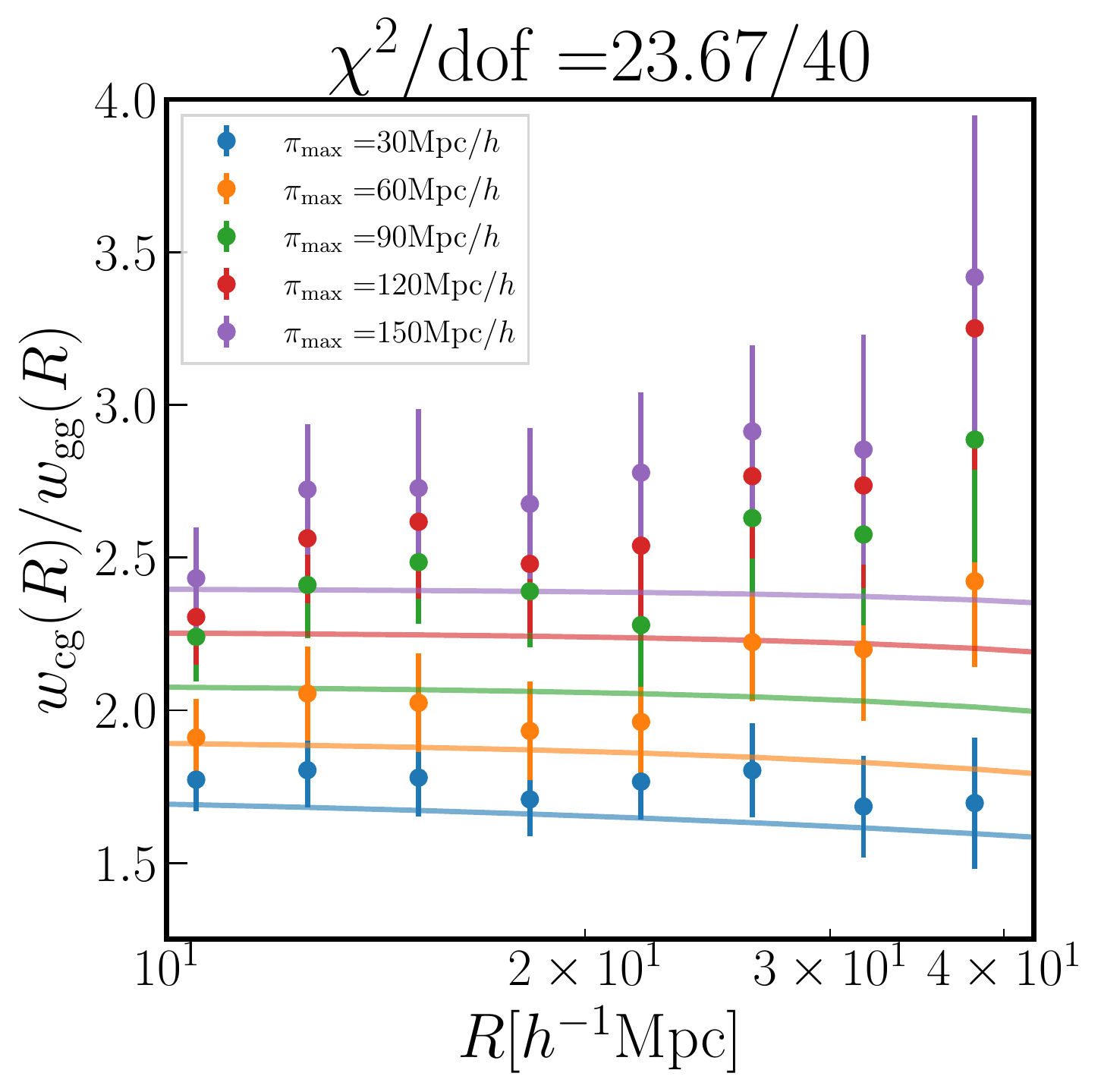}
    \caption{\label{fig:obs_pred_30to40}
   The projected cross-correlation functions of SDSS RM clusters and galaxies relative to the galaxy auto-correlation functions with $\pi_{\rm max}=30, 60, 90, 120$, and $150h^{-1}{\rm Mpc}$ with $30 \leq \lambda \leq 40$. The circles with error bars are from the measurements and the solid lines are the predictions based on our model with the best-fit parameters.
    }
\end{figure}

Lastly, Fig.~\ref{fig:alphas_obs} shows the best-fit value for the anisotropic boost $\alpha$ as a function of richness $\lambda$.
Unlike the result from the simulation analysis whose $\alpha$ values decrease as richness $\lambda$ increases, the $\alpha$ values are almost constant ($\sim 20\%$) across all the richness bins except the one with $30 \leq \lambda \leq 40$.
The value of $\alpha$ for the cluster sample with $30 \leq \lambda \leq 40$ is $55 \pm 15 \%$, which is unexpectedly large. 
To investigate further the cause of this large $\alpha$ value, Fig.~\ref{fig:obs_pred_30to40} shows the measured ratio $w_{\rm cg}(R)/w_{\rm gg}(R)$ and the prediction from the best-fit parameters for the case of $30 \leq \lambda \leq 40$.
Unlike Fig.~\ref{fig:obs_pred_20to30}, the best-fit model does not fit well with the measured ratio, in particular for the case of large $\pi_{\rm max}$.
While the ratio with $\pi_{\rm max}=30h^{-1}{\rm Mpc}$ is almost constant on $10h^{-1}{\rm Mpc} \leq R \leq 42 h^{-1}{\rm Mpc}$, the ratios with other $\pi_{\rm max}$ increase as $R$ increases.
This increase is against the predicted scale-dependence by the RSD effect, and it is unclear what can cause this increase in the ratio.
We tried different $R$ ranges to constrain the value of $\alpha$. However, any choice of $R$ did not significantly change the best-fit value of $\alpha$.
We will leave the further investigation of the cause to the future work.

\section{Summary}
\label{sec:summary}
In this paper, we implemented the model and the method to quantitatively evaluate the anisotropic boost to the cluster clustering and lensing due to the projection effects. We validated our model against the mock cluster catalogs and then applied it to the SDSS redMaPPer cluster catalog. We summarize our conclusions as follows:
\begin{itemize}
    \item The anisotropic boosts can be quantitatively measured using the cluster-galaxy projected correlation functions concerning galaxy auto-correlation functions. We assume that the selection of galaxies is isotropic, and varying integral scales is a key to measure the boost. 
    \item To model the projection effects, we made two assumptions: $\alpha$ increases constantly as $\pi_{\rm max}$ increases, and the increase of $\alpha$ stops when $\pi_{\rm max}>d_{\rm proj}$.
    \item Upon validation with the mock cluster catalog using the projection length of $d_{\rm proj}=120h^{-1}{\rm Mpc}$, our model was able to measure the expected anisotropic boost through the cluster-galaxy cross-correlation functions.
    \item We applied our model to the SDSS redMaPPer cluster catalog and measured the boost factor to be roughly $\sim 20\%$ for all the richness bins except the $\lambda \in [30,40]$ bin. The size of the boosts is consistent with the constraints in the cluster cosmology analysis by \cite{Park_etal2021} and \cite{To_Krause2021}. 
    \item While the ratios $w_{\rm cg}(R)/w_{\rm gg}(R)$ of the mock cluster samples exhibit the scale-dependence consistent with the prediction from the RSD effect, the ratios measured from the SDSS redMaPPer clusters show little or opposite scale-dependence. Understanding the cause of this is our future work.
    \item Our model also enabled to constrain the projection length of clusters along the LOS. Even though the accuracy is somewhat questionable based on the validation from the mock, our model constrained $d_{\rm proj}$ to be $ \geq 100h^{-1}{\rm Mpc}$.
\end{itemize}

In this work, we develop a method to measure the anisotropic boost on the amplitude of cluster clustering through cluster-galaxy cross-correlation functions and measure the anisotropic boost in the SDSS redMaPPer clusters is roughly $20\%$ except the clusters with $30  \leq  \lambda  \leq  40$. We plan to investigate further the cause of large anisotropic boost ($\sim 60\%$) for these clusters in our future work. Additionally, we plan to extend our method for clusters without spectroscopic follow-ups in our future works.

\section*{Acknowledgements}

We thank Youngsoo Park, Peter Behroozi, Eduardo Rozo, Ryuichi Takahashi and Kiyotomo Ichiki for their comments, and thank Surhud More, Takahiro Nishimichi and Dark Quest collaboration team who provided an original cluster finder and Dark Quest simulations. TS thanks Masahiro Takada to encourages this publication.
TS is supported by Grant-in-Aid for JSPS Fellows 20J01600 and JSPS KAKENHI Grant Number 20H05855.

Funding for SDSS-III has been provided by the Alfred P. Sloan Foundation, the Participating Institutions, the National Science Foundation, and the U.S. Department of Energy Office of Science. The SDSS-III web site is \url{http://www.sdss3.org/.}

SDSS-III is managed by the Astrophysical Research Consortium for the Participating Institutions of the SDSS-III Collaboration including the University of Arizona, the Brazilian Participation Group, Brookhaven National Laboratory, Carnegie Mellon University, University of Florida, the French Participation Group, the German Participation Group, Harvard University, the Instituto de Astrofisica de Canarias, the Michigan State/Notre Dame/JINA Participation Group, Johns Hopkins University, Lawrence Berkeley National Laboratory, Max Planck Institute for Astrophysics, Max Planck Institute for Extraterrestrial Physics, New Mexico State University, New York University, Ohio State University, Pennsylvania State University, University of Portsmouth, Princeton University, the Spanish Participation Group, University of Tokyo, University of Utah, Vanderbilt University, University of Virginia, University of Washington, and Yale University.




\bibliographystyle{mnras}
\bibliography{refs} 

\begin{thebibliography}{}
\makeatletter
\relax
\def\mn@urlcharsother{\let\do\@makeother \do\$\do\&\do\#\do\^\do\_\do\%\do\~}
\def\mn@doi{\begingroup\mn@urlcharsother \@ifnextchar [ {\mn@doi@}
  {\mn@doi@[]}}
\def\mn@doi@[#1]#2{\def\@tempa{#1}\ifx\@tempa\@empty \href
  {http://dx.doi.org/#2} {doi:#2}\else \href {http://dx.doi.org/#2} {#1}\fi
  \endgroup}
\def\mn@eprint#1#2{\mn@eprint@#1:#2::\@nil}
\def\mn@eprint@arXiv#1{\href {http://arxiv.org/abs/#1} {{\tt arXiv:#1}}}
\def\mn@eprint@dblp#1{\href {http://dblp.uni-trier.de/rec/bibtex/#1.xml}
  {dblp:#1}}
\def\mn@eprint@#1:#2:#3:#4\@nil{\def\@tempa {#1}\def\@tempb {#2}\def\@tempc
  {#3}\ifx \@tempc \@empty \let \@tempc \@tempb \let \@tempb \@tempa \fi \ifx
  \@tempb \@empty \def\@tempb {arXiv}\fi \@ifundefined
  {mn@eprint@\@tempb}{\@tempb:\@tempc}{\expandafter \expandafter \csname
  mn@eprint@\@tempb\endcsname \expandafter{\@tempc}}}

\bibitem[\protect\citeauthoryear{{Aihara} et~al.,}{{Aihara}
  et~al.}{2011}]{Aihara:2011}
{Aihara} H.,  et~al., 2011, \mn@doi [\apjs] {10.1088/0067-0049/193/2/29}, \href
  {http://adsabs.harvard.edu/abs/2011ApJS..193...29A} {193, 29}

\bibitem[\protect\citeauthoryear{{Aihara} et~al.,}{{Aihara}
  et~al.}{2018}]{HSCOverview:17}
{Aihara} H.,  et~al., 2018, \mn@doi [\pasj] {10.1093/pasj/psx066}, \href
  {https://ui.adsabs.harvard.edu/abs/2018PASJ...70S...4A} {70, S4}

\bibitem[\protect\citeauthoryear{{Alam} et~al.,}{{Alam}
  et~al.}{2015}]{Alam:2015}
{Alam} S.,  et~al., 2015, \mn@doi [\apjs] {10.1088/0067-0049/219/1/12}, \href
  {http://adsabs.harvard.edu/abs/2015ApJS..219...12A} {219, 12}

\bibitem[\protect\citeauthoryear{{Amendola} et~al.,}{{Amendola}
  et~al.}{2018}]{euclid2018}
{Amendola} L.,  et~al., 2018, \mn@doi [Living Reviews in Relativity]
  {10.1007/s41114-017-0010-3}, \href
  {https://ui.adsabs.harvard.edu/abs/2018LRR....21....2A} {21, 2}

\bibitem[\protect\citeauthoryear{{Bartelmann}}{{Bartelmann}}{1996}]{Bartelmann96}
{Bartelmann} M.,  1996, \aap, \href
  {http://ads.nao.ac.jp/abs/1996A%26A...313..697B} {313, 697}

\bibitem[\protect\citeauthoryear{{Behroozi}, {Wechsler}  \& {Wu}}{{Behroozi}
  et~al.}{2013}]{Behroozi:2013}
{Behroozi} P.~S.,  {Wechsler} R.~H.,   {Wu} H.-Y.,  2013, \mn@doi [\apj]
  {10.1088/0004-637X/762/2/109}, \href
  {http://adsabs.harvard.edu/abs/2013ApJ...762..109B} {762, 109}

\bibitem[\protect\citeauthoryear{{Behroozi}, {Wechsler}, {Hearin}  \&
  {Conroy}}{{Behroozi} et~al.}{2019}]{Behroozi2019}
{Behroozi} P.,  {Wechsler} R.~H.,  {Hearin} A.~P.,   {Conroy} C.,  2019,
  \mn@doi [\mnras] {10.1093/mnras/stz1182}, \href
  {https://ui.adsabs.harvard.edu/abs/2019MNRAS.488.3143B} {488, 3143}

\bibitem[\protect\citeauthoryear{{Busch} \& {White}}{{Busch} \&
  {White}}{2017}]{BuschWhite:17}
{Busch} P.,  {White} S.~D.~M.,  2017, \mn@doi [\mnras] {10.1093/mnras/stx1584},
  \href {http://adsabs.harvard.edu/abs/2017MNRAS.470.4767B} {470, 4767}

\bibitem[\protect\citeauthoryear{{Costanzi} et~al.,}{{Costanzi}
  et~al.}{2019a}]{costanzietalprojection}
{Costanzi} M.,  et~al., 2019a, \mn@doi [\mnras] {10.1093/mnras/sty2665}, \href
  {https://ui.adsabs.harvard.edu/abs/2019MNRAS.482..490C} {482, 490}

\bibitem[\protect\citeauthoryear{{Costanzi} et~al.,}{{Costanzi}
  et~al.}{2019b}]{2018arXiv181009456C}
{Costanzi} M.,  et~al., 2019b, \mn@doi [\mnras] {10.1093/mnras/stz1949}, \href
  {https://ui.adsabs.harvard.edu/abs/2019MNRAS.488.4779C} {488, 4779}

\bibitem[\protect\citeauthoryear{{Crocce} \& {Scoccimarro}}{{Crocce} \&
  {Scoccimarro}}{2006}]{crocce06b}
{Crocce} M.,  {Scoccimarro} R.,  2006, \mn@doi [\prd]
  {10.1103/PhysRevD.73.063519}, \href
  {http://adsabs.harvard.edu/abs/2006PhRvD..73f3519C} {73, 063519}

\bibitem[\protect\citeauthoryear{{Crocce}, {Pueblas}  \&
  {Scoccimarro}}{{Crocce} et~al.}{2006}]{crocce06a}
{Crocce} M.,  {Pueblas} S.,   {Scoccimarro} R.,  2006, \mn@doi [Mon. Not. Roy.
  Astron. Soc.] {10.1111/j.1365-2966.2006.11040.x}, \href
  {http://ads.nao.ac.jp/abs/2006MNRAS.373..369C} {373, 369}

\bibitem[\protect\citeauthoryear{{Diemer} \& {Kravtsov}}{{Diemer} \&
  {Kravtsov}}{2015}]{2015ApJ...799..108D}
{Diemer} B.,  {Kravtsov} A.~V.,  2015, \mn@doi [\apj]
  {10.1088/0004-637X/799/1/108}, \href
  {https://ui.adsabs.harvard.edu/abs/2015ApJ...799..108D} {799, 108}

\bibitem[\protect\citeauthoryear{{Dore} et~al.,}{{Dore}
  et~al.}{2019}]{WFIRST2019}
{Dore} O.,  et~al., 2019, \baas, \href
  {https://ui.adsabs.harvard.edu/abs/2019BAAS...51c.341D} {51, 341}

\bibitem[\protect\citeauthoryear{Foreman-Mackey, Hogg, Lang  \&
  Goodman}{Foreman-Mackey et~al.}{2013}]{emcee}
Foreman-Mackey D.,  Hogg D.~W.,  Lang D.,   Goodman J.,  2013, \mn@doi
  [Publications of the Astronomical Society of the Pacific] {10.1086/670067},
  125, 306–312

\bibitem[\protect\citeauthoryear{{Haiman}, {Mohr}  \& {Holder}}{{Haiman}
  et~al.}{2001}]{Haiman:2001}
{Haiman} Z.,  {Mohr} J.~J.,   {Holder} G.~P.,  2001, \mn@doi [\apj]
  {10.1086/320939}, \href {http://adsabs.harvard.edu/abs/2001ApJ...553..545H}
  {553, 545}

\bibitem[\protect\citeauthoryear{{Hartlap}, {Simon}  \& {Schneider}}{{Hartlap}
  et~al.}{2007}]{Hartlapetal2007}
{Hartlap} J.,  {Simon} P.,   {Schneider} P.,  2007, \mn@doi [\aap]
  {10.1051/0004-6361:20066170}, \href
  {http://adsabs.harvard.edu/abs/2007A%26A...464..399H} {464, 399}

\bibitem[\protect\citeauthoryear{{Jing}, {Mo}  \& {B{\"o}rner}}{{Jing}
  et~al.}{1998}]{1998ApJ...494....1J}
{Jing} Y.~P.,  {Mo} H.~J.,   {B{\"o}rner} G.,  1998, \mn@doi [\apj]
  {10.1086/305209}, \href {http://adsabs.harvard.edu/abs/1998ApJ...494....1J}
  {494, 1}

\bibitem[\protect\citeauthoryear{Kobayashi, Nishimichi, Takada  \&
  Takahashi}{Kobayashi et~al.}{2020}]{Kobayashi:2019jrn}
Kobayashi Y.,  Nishimichi T.,  Takada M.,   Takahashi R.,  2020, \mn@doi [Phys.
  Rev.] {10.1103/PhysRevD.101.023510}, D101, 023510

\bibitem[\protect\citeauthoryear{{Kuijken} et~al.,}{{Kuijken}
  et~al.}{2015}]{KiDs2015}
{Kuijken} K.,  et~al., 2015, \mn@doi [\mnras] {10.1093/mnras/stv2140}, \href
  {https://ui.adsabs.harvard.edu/abs/2015MNRAS.454.3500K} {454, 3500}

\bibitem[\protect\citeauthoryear{{LSST Science Collaboration} et~al.,}{{LSST
  Science Collaboration} et~al.}{2009}]{LSST2009}
{LSST Science Collaboration} et~al., 2009, arXiv e-prints, \href
  {https://ui.adsabs.harvard.edu/abs/2009arXiv0912.0201L} {p. arXiv:0912.0201}

\bibitem[\protect\citeauthoryear{{Landy} \& {Szalay}}{{Landy} \&
  {Szalay}}{1993}]{LandySzalay:93}
{Landy} S.~D.,  {Szalay} A.~S.,  1993, \mn@doi [\apj] {10.1086/172900}, \href
  {http://adsabs.harvard.edu/abs/1993ApJ...412...64L} {412, 64}

\bibitem[\protect\citeauthoryear{{Lewis}, {Challinor}  \& {Lasenby}}{{Lewis}
  et~al.}{2000}]{camb}
{Lewis} A.,  {Challinor} A.,   {Lasenby} A.,  2000, \mn@doi [Astrophys. J.]
  {10.1086/309179}, \href {http://ads.nao.ac.jp/abs/2000ApJ...538..473L} {538,
  473}

\bibitem[\protect\citeauthoryear{{Miyatake}, {More}, {Takada}, {Spergel},
  {Mandelbaum}, {Rykoff}  \& {Rozo}}{{Miyatake} et~al.}{2016}]{Miyatake:2016}
{Miyatake} H.,  {More} S.,  {Takada} M.,  {Spergel} D.~N.,  {Mandelbaum} R.,
  {Rykoff} E.~S.,   {Rozo} E.,  2016, \mn@doi [Physical Review Letters]
  {10.1103/PhysRevLett.116.041301}, \href
  {http://adsabs.harvard.edu/abs/2016PhRvL.116d1301M} {116, 041301}

\bibitem[\protect\citeauthoryear{{Navarro}, {Frenk}  \& {White}}{{Navarro}
  et~al.}{1997}]{nfw97}
{Navarro} J.~F.,  {Frenk} C.~S.,   {White} S.~D.~M.,  1997, \mn@doi [\apj]
  {10.1086/304888}, \href {http://adsabs.harvard.edu/abs/1997ApJ...490..493N}
  {490, 493}

\bibitem[\protect\citeauthoryear{{Nishimichi} et~al.,}{{Nishimichi}
  et~al.}{2009}]{nishimichi09}
{Nishimichi} T.,  et~al., 2009, Publ. Astron. Soc. Japan, \href
  {http://adsabs.harvard.edu/abs/2009PASJ...61..321N} {61, 321}

\bibitem[\protect\citeauthoryear{{Nishimichi} et~al.,}{{Nishimichi}
  et~al.}{2019}]{darkemu}
{Nishimichi} T.,  et~al., 2019, \mn@doi [\apj] {10.3847/1538-4357/ab3719},
  \href {https://ui.adsabs.harvard.edu/abs/2019ApJ...884...29N} {884, 29}

\bibitem[\protect\citeauthoryear{{Oguri} \& {Takada}}{{Oguri} \&
  {Takada}}{2011}]{OguriTakada:2011}
{Oguri} M.,  {Takada} M.,  2011, \mn@doi [\prd] {10.1103/PhysRevD.83.023008},
  \href {http://adsabs.harvard.edu/abs/2011PhRvD..83b3008O} {83, 023008}

\bibitem[\protect\citeauthoryear{Parejko et~al.,}{Parejko
  et~al.}{2012}]{parejko_etal2012}
Parejko J.~K.,  et~al., 2012, \mn@doi [Monthly Notices of the Royal
  Astronomical Society] {10.1093/mnras/sts314}, 429, 98–112

\bibitem[\protect\citeauthoryear{{Park}, {Sunayama}, {Takada}, {Kobayashi},
  {Miyatake}, {More}, {Nishimichi}  \& {Sugiyama}}{{Park}
  et~al.}{2021}]{Park_etal2021}
{Park} Y.,  {Sunayama} T.,  {Takada} M.,  {Kobayashi} Y.,  {Miyatake} H.,
  {More} S.,  {Nishimichi} T.,   {Sugiyama} S.,  2021, arXiv e-prints, \href
  {https://ui.adsabs.harvard.edu/abs/2021arXiv211209059P} {p. arXiv:2112.09059}

\bibitem[\protect\citeauthoryear{{Peacock} \& {Smith}}{{Peacock} \&
  {Smith}}{2000}]{peacock:2000qy}
{Peacock} J.~A.,  {Smith} R.~E.,  2000, \mn@doi [\mnras]
  {10.1046/j.1365-8711.2000.03779.x}, \href
  {http://adsabs.harvard.edu/abs/2000MNRAS.318.1144P} {318, 1144}

\bibitem[\protect\citeauthoryear{{Planck Collaboration} et~al.,}{{Planck
  Collaboration} et~al.}{2016}]{Planck:2015}
{Planck Collaboration} et~al., 2016, \mn@doi [\aap]
  {10.1051/0004-6361/201525830}, \href
  {http://adsabs.harvard.edu/abs/2016A%26A...594A..13P} {594, A13}

\bibitem[\protect\citeauthoryear{{Rozo} \& {Rykoff}}{{Rozo} \&
  {Rykoff}}{2014}]{Rozo2014}
{Rozo} E.,  {Rykoff} E.~S.,  2014, \mn@doi [\apj] {10.1088/0004-637X/783/2/80},
  \href {http://ads.nao.ac.jp/abs/2014ApJ...783...80R} {783, 80}

\bibitem[\protect\citeauthoryear{{Rozo} et~al.,}{{Rozo}
  et~al.}{2010}]{Rozoetal:10}
{Rozo} E.,  et~al., 2010, \mn@doi [\apj] {10.1088/0004-637X/708/1/645}, \href
  {http://adsabs.harvard.edu/abs/2010ApJ...708..645R} {708, 645}

\bibitem[\protect\citeauthoryear{{Rozo}, {Rykoff}, {Bartlett}  \&
  {Melin}}{{Rozo} et~al.}{2015a}]{Rozo2015}
{Rozo} E.,  {Rykoff} E.~S.,  {Bartlett} J.~G.,   {Melin} J.-B.,  2015a, \mn@doi
  [\mnras] {10.1093/mnras/stv605}, \href
  {http://ads.nao.ac.jp/abs/2015MNRAS.450..592R} {450, 592}

\bibitem[\protect\citeauthoryear{{Rozo}, {Rykoff}, {Becker}, {Reddick}  \&
  {Wechsler}}{{Rozo} et~al.}{2015b}]{Rozo2015_2}
{Rozo} E.,  {Rykoff} E.~S.,  {Becker} M.,  {Reddick} R.~M.,   {Wechsler} R.~H.,
   2015b, \mn@doi [\mnras] {10.1093/mnras/stv1560}, \href
  {http://ads.nao.ac.jp/abs/2015MNRAS.453...38R} {453, 38}

\bibitem[\protect\citeauthoryear{{Rykoff} et~al.,}{{Rykoff}
  et~al.}{2014}]{Rykoff_etal2014}
{Rykoff} E.~S.,  et~al., 2014, \mn@doi [\apj] {10.1088/0004-637X/785/2/104},
  \href {http://ads.nao.ac.jp/abs/2014ApJ...785..104R} {785, 104}

\bibitem[\protect\citeauthoryear{{Scoccimarro}}{{Scoccimarro}}{1998}]{scoccimarro98}
{Scoccimarro} R.,  1998, \mn@doi [Mon. Not. Roy. Astron. Soc.]
  {10.1046/j.1365-8711.1998.01845.x}, \href
  {http://adsabs.harvard.edu/abs/1998MNRAS.299.1097S} {299, 1097}

\bibitem[\protect\citeauthoryear{{Seljak}}{{Seljak}}{2000}]{seljak:2000uq}
{Seljak} U.,  2000, \mn@doi [\mnras] {10.1046/j.1365-8711.2000.03715.x}, \href
  {http://adsabs.harvard.edu/abs/2000MNRAS.318..203S} {318, 203}

\bibitem[\protect\citeauthoryear{{Springel}}{{Springel}}{2005}]{Springel:2005}
{Springel} V.,  2005, \mn@doi [\mnras] {10.1111/j.1365-2966.2005.09655.x},
  \href {http://adsabs.harvard.edu/abs/2005MNRAS.364.1105S} {364, 1105}

\bibitem[\protect\citeauthoryear{{Sunayama} \& {More}}{{Sunayama} \&
  {More}}{2019}]{SunayamaMore}
{Sunayama} T.,  {More} S.,  2019, \mn@doi [\mnras] {10.1093/mnras/stz2832},
  \href {https://ui.adsabs.harvard.edu/abs/2019MNRAS.490.4945S} {490, 4945}

\bibitem[\protect\citeauthoryear{{Sunayama} et~al.,}{{Sunayama}
  et~al.}{2020}]{Sunayama_etal2020}
{Sunayama} T.,  et~al., 2020, arXiv e-prints, \href
  {https://ui.adsabs.harvard.edu/abs/2020arXiv200203867S} {p. arXiv:2002.03867}

\bibitem[\protect\citeauthoryear{{Takada} \& {Bridle}}{{Takada} \&
  {Bridle}}{2007}]{TakadaBridle:07}
{Takada} M.,  {Bridle} S.,  2007, \mn@doi [New Journal of Physics]
  {10.1088/1367-2630/9/12/446}, \href
  {http://adsabs.harvard.edu/abs/2007NJPh....9..446T} {9, 446}

\bibitem[\protect\citeauthoryear{{The Dark Energy Survey Collaboration}}{{The
  Dark Energy Survey Collaboration}}{2005}]{DES2005}
{The Dark Energy Survey Collaboration} 2005, arXiv e-prints, \href
  {https://ui.adsabs.harvard.edu/abs/2005astro.ph.10346T} {pp
  astro--ph/0510346}

\bibitem[\protect\citeauthoryear{To et~al.,}{To et~al.}{2021}]{To_Krause2021}
To C.,  et~al., 2021, \mn@doi [Physical Review Letters]
  {10.1103/physrevlett.126.141301}, 126

\bibitem[\protect\citeauthoryear{{Valageas} \& {Nishimichi}}{{Valageas} \&
  {Nishimichi}}{2011}]{Valageas_Nishimichi2011}
{Valageas} P.,  {Nishimichi} T.,  2011, \mn@doi [\aap]
  {10.1051/0004-6361/201015685}, \href
  {https://ui.adsabs.harvard.edu/abs/2011A&A...527A..87V} {527, A87}

\bibitem[\protect\citeauthoryear{{Vikhlinin} et~al.,}{{Vikhlinin}
  et~al.}{2009}]{Vikhlininetal:09}
{Vikhlinin} A.,  et~al., 2009, \mn@doi [\apj] {10.1088/0004-637X/692/2/1060},
  \href {http://adsabs.harvard.edu/abs/2009ApJ...692.1060V} {692, 1060}

\bibitem[\protect\citeauthoryear{{Weinberg}, {Mortonson}, {Eisenstein},
  {Hirata}, {Riess}  \& {Rozo}}{{Weinberg} et~al.}{2013}]{Weinberg:2013}
{Weinberg} D.~H.,  {Mortonson} M.~J.,  {Eisenstein} D.~J.,  {Hirata} C.,
  {Riess} A.~G.,   {Rozo} E.,  2013, \mn@doi [\physrep]
  {10.1016/j.physrep.2013.05.001}, \href
  {http://adsabs.harvard.edu/abs/2013PhR...530...87W} {530, 87}

\bibitem[\protect\citeauthoryear{{White}, {Efstathiou}  \& {Frenk}}{{White}
  et~al.}{1993}]{Whieetal:93}
{White} S.~D.~M.,  {Efstathiou} G.,   {Frenk} C.~S.,  1993, \mn@doi [\mnras]
  {10.1093/mnras/262.4.1023}, \href
  {http://adsabs.harvard.edu/abs/1993MNRAS.262.1023W} {262, 1023}

\bibitem[\protect\citeauthoryear{{Zheng} et~al.,}{{Zheng}
  et~al.}{2005}]{2005ApJ...633..791Z}
{Zheng} Z.,  et~al., 2005, \mn@doi [\apj] {10.1086/466510}, \href
  {http://adsabs.harvard.edu/abs/2005ApJ...633..791Z} {633, 791}

\bibitem[\protect\citeauthoryear{van~den Bosch, More, Cacciato, Mo  \&
  Yang}{van~den Bosch et~al.}{2013}]{vdBosch2013}
van~den Bosch F.~C.,  More S.,  Cacciato M.,  Mo H.,   Yang X.,  2013, \mn@doi
  [Monthly Notices of the Royal Astronomical Society] {10.1093/mnras/sts006},
  430, 725–746

\makeatother
\end{thebibliography}




\appendix

\section{Boosts measured from lensing signals for all richness bins and the projection lengths}
\label{sec:appA}

\begin{figure*}
    \centering
    \includegraphics[width=0.3\textwidth]{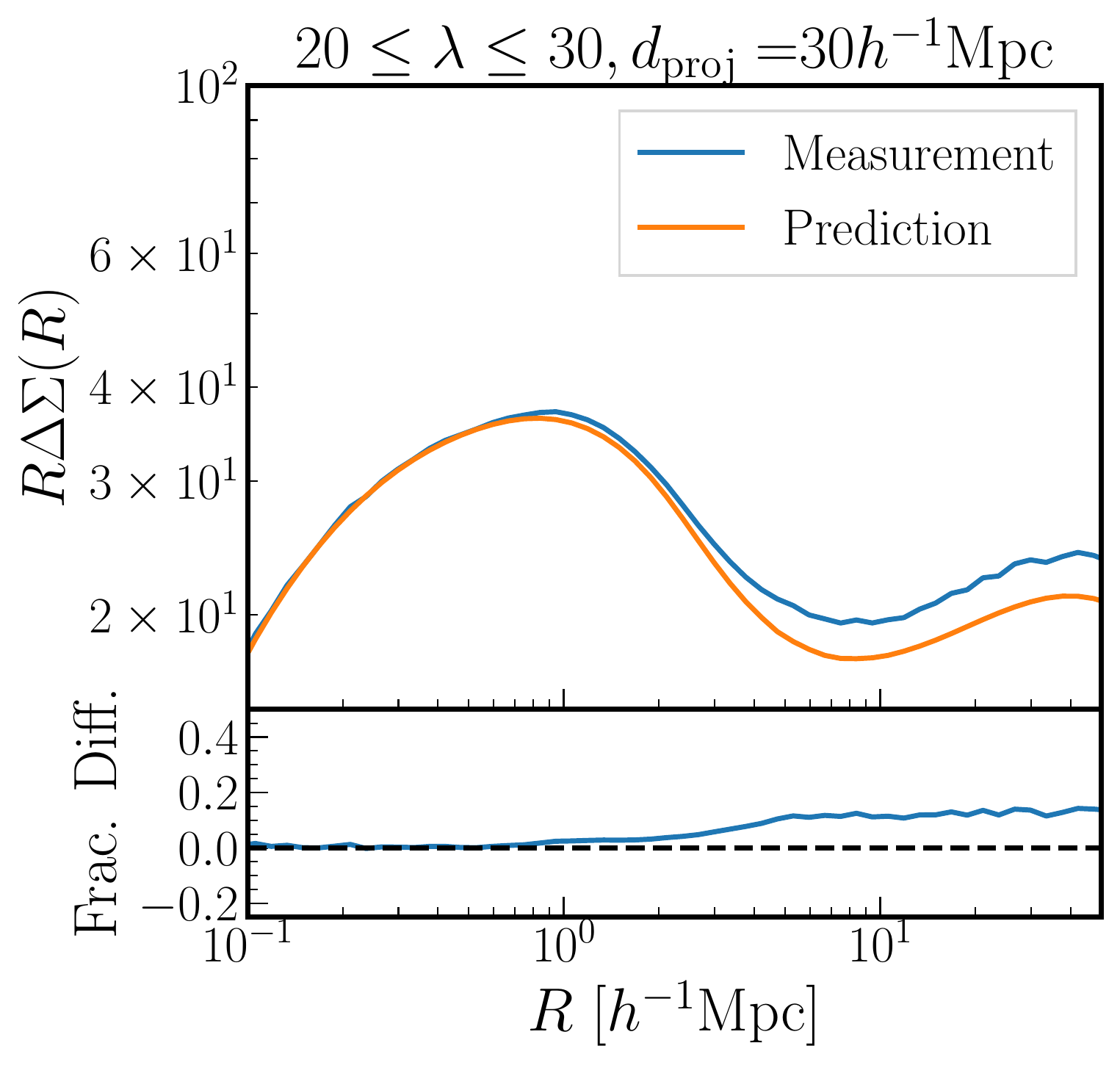}
    \includegraphics[width=0.3\textwidth]{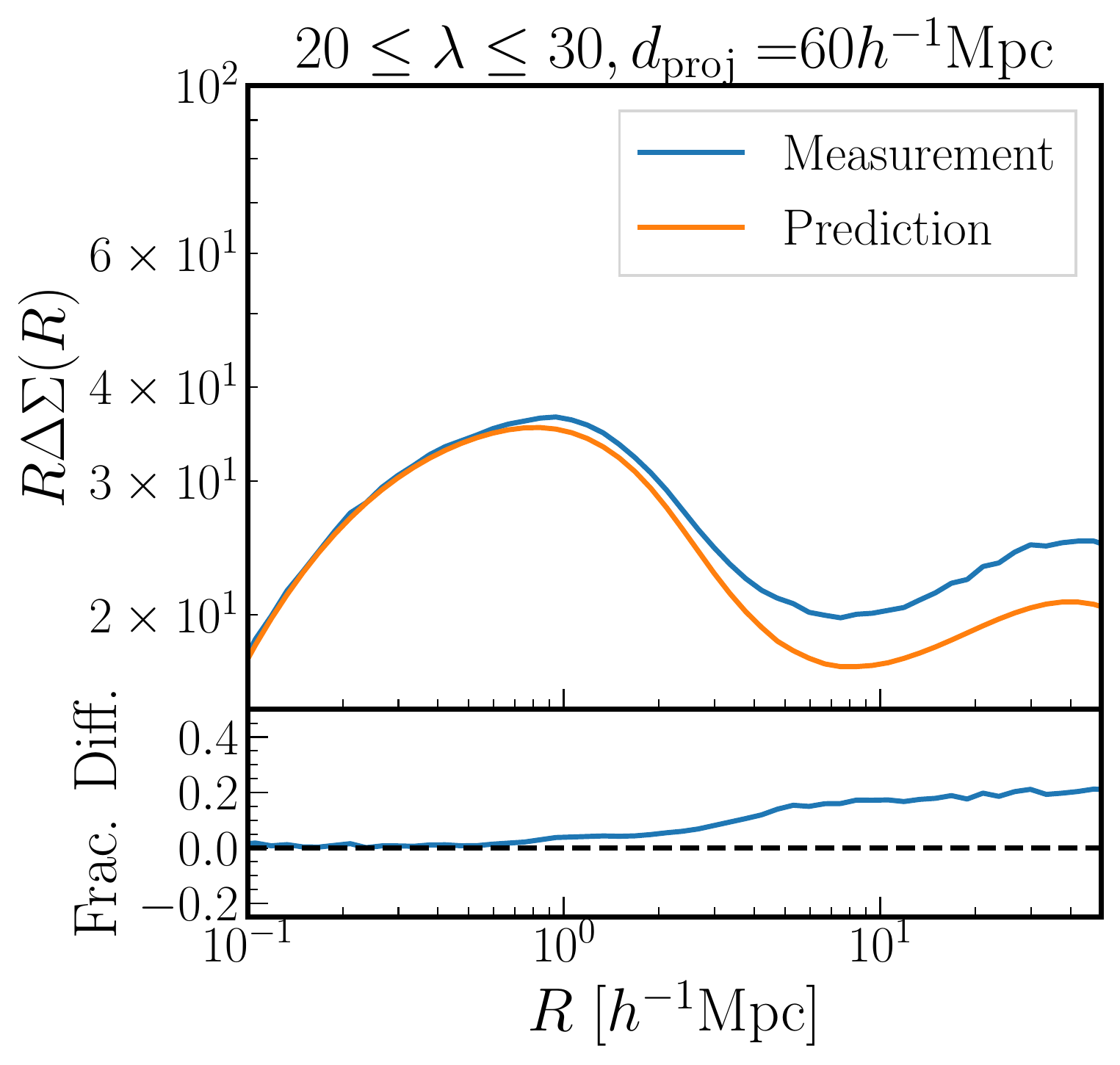}
    \includegraphics[width=0.3\textwidth]{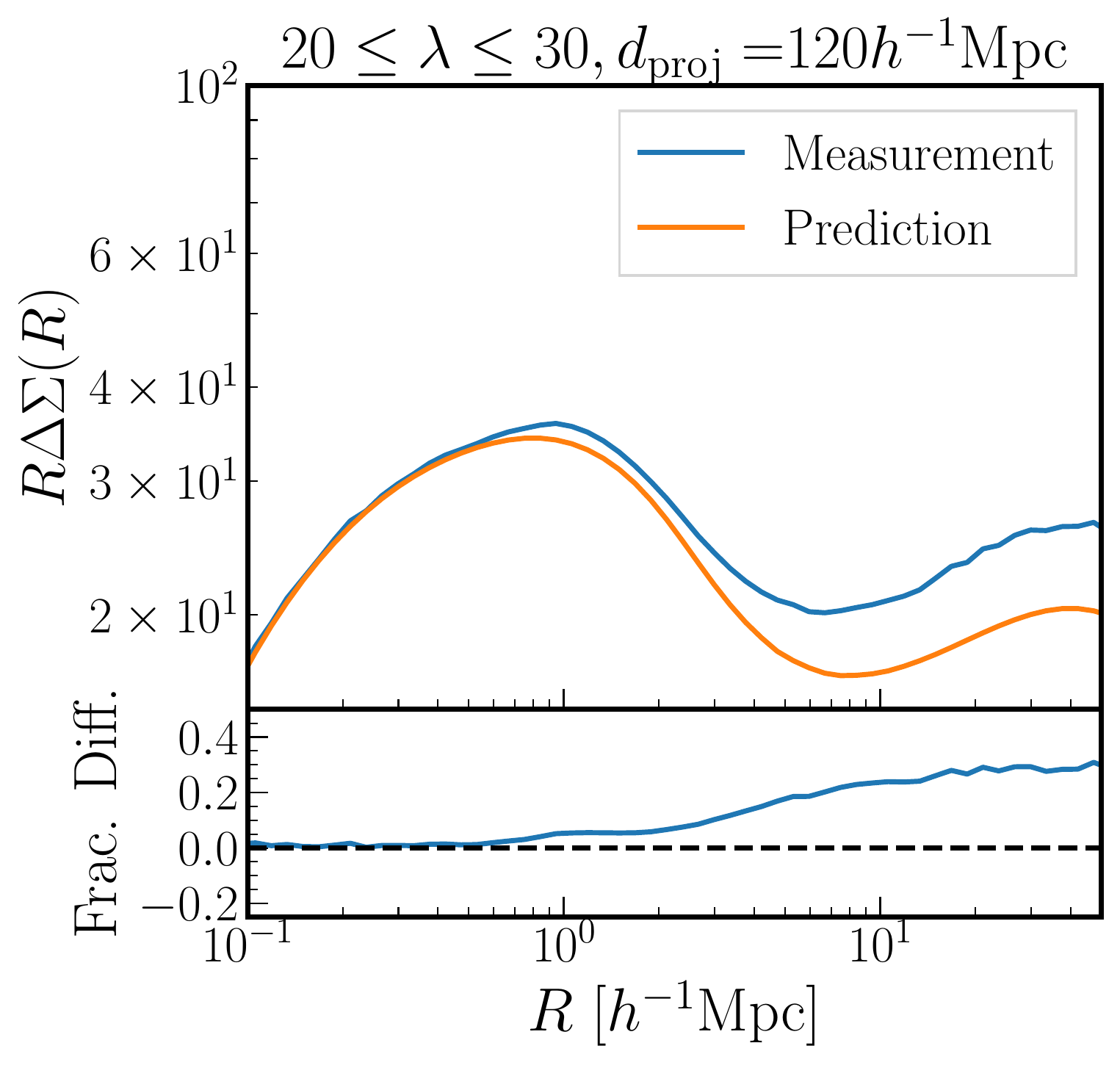}
    
    \vspace{0.3cm}
    
    \includegraphics[width=0.3\textwidth]{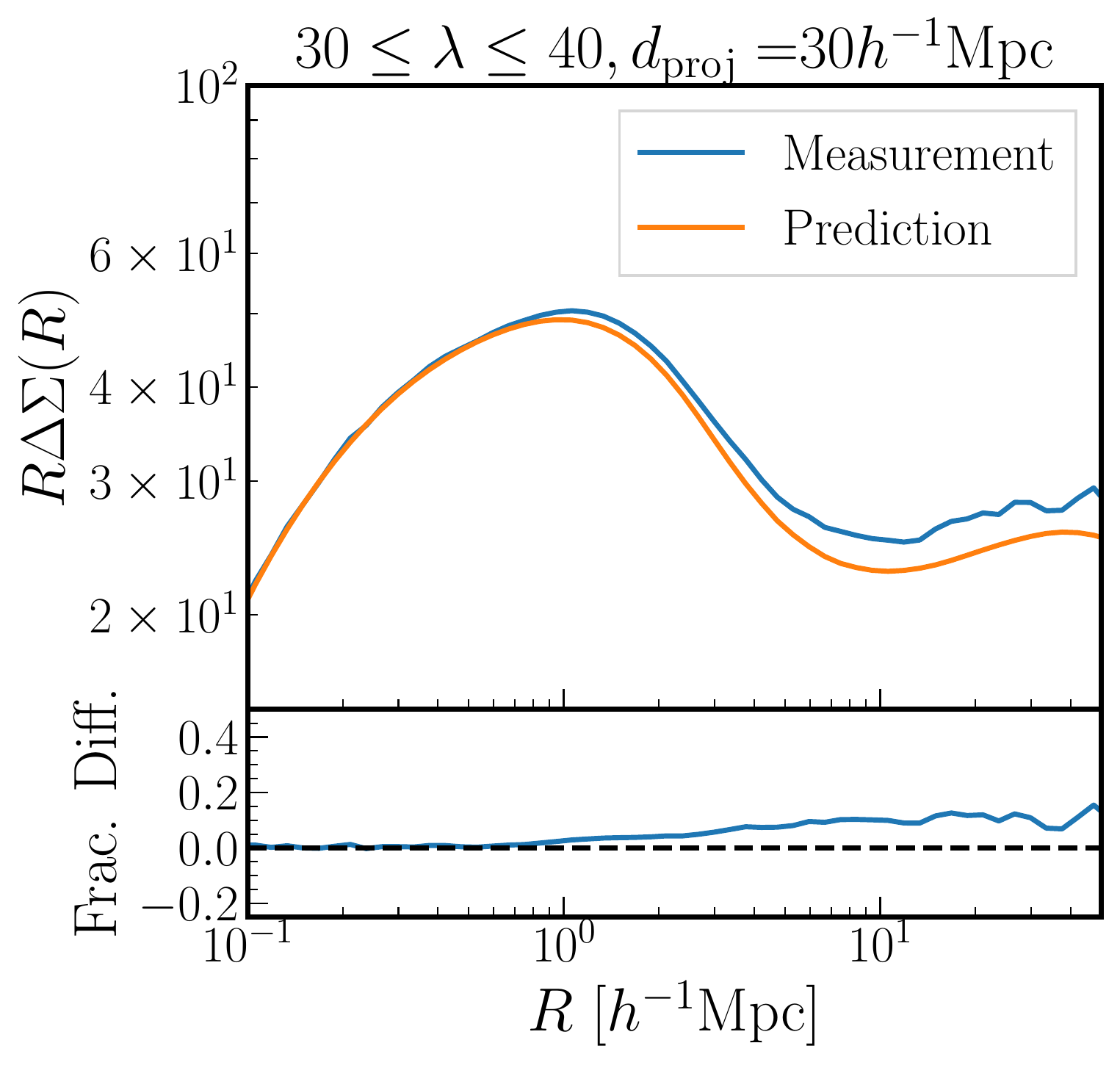}
    \includegraphics[width=0.3\textwidth]{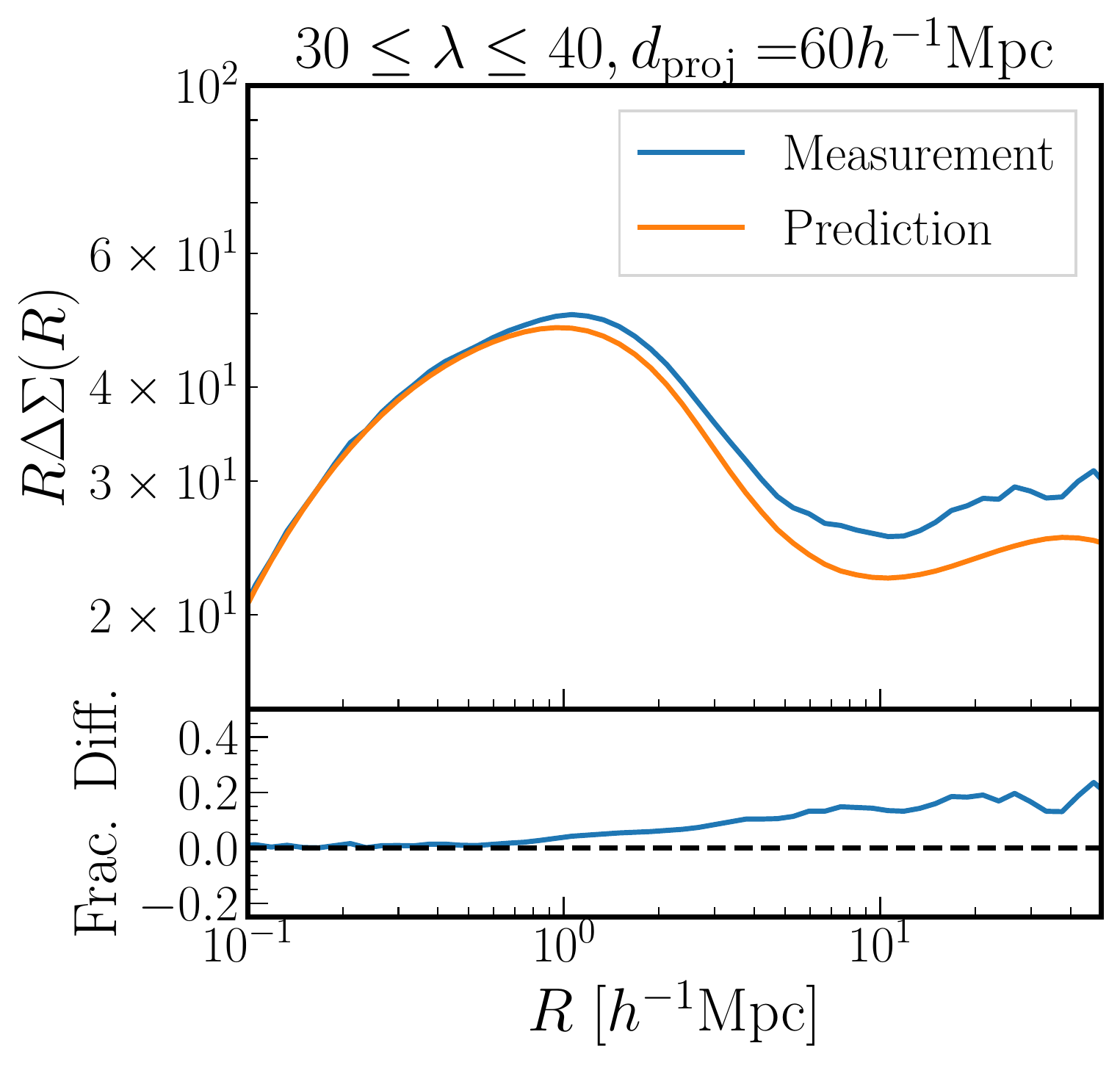}
    \includegraphics[width=0.3\textwidth]{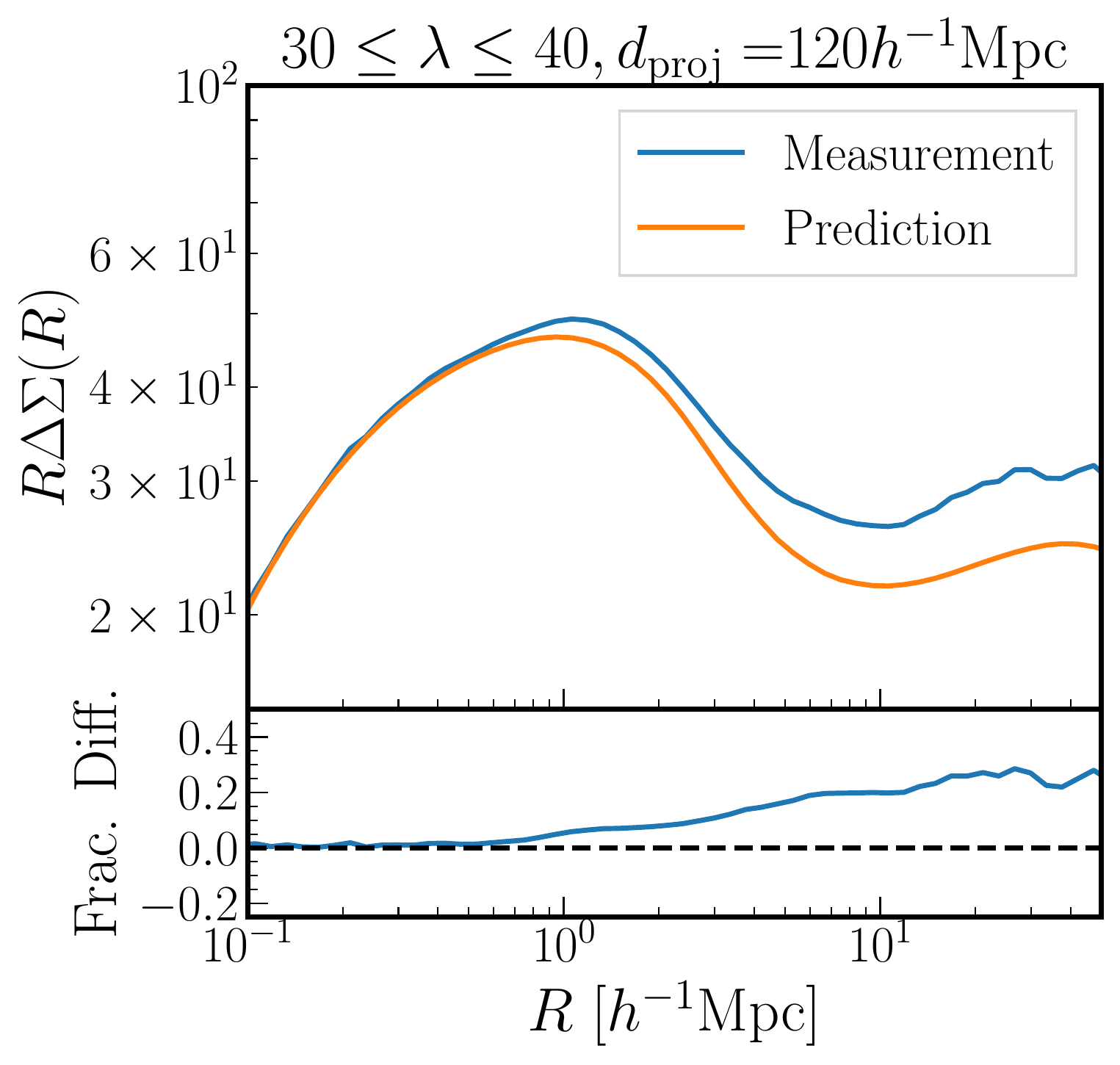}
    
    \vspace{0.3cm}
    \includegraphics[width=0.3\textwidth]{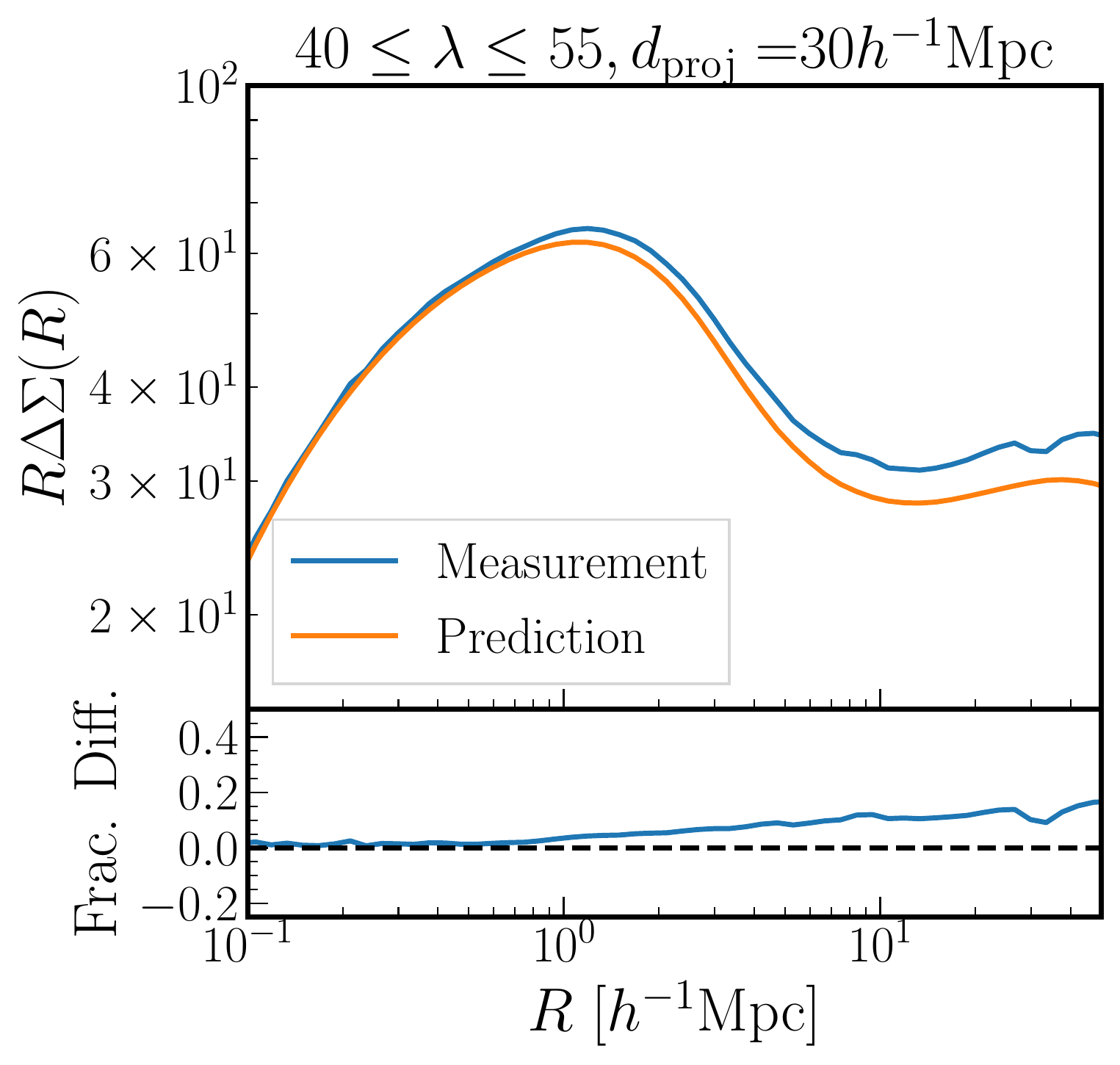}
    \includegraphics[width=0.3\textwidth]{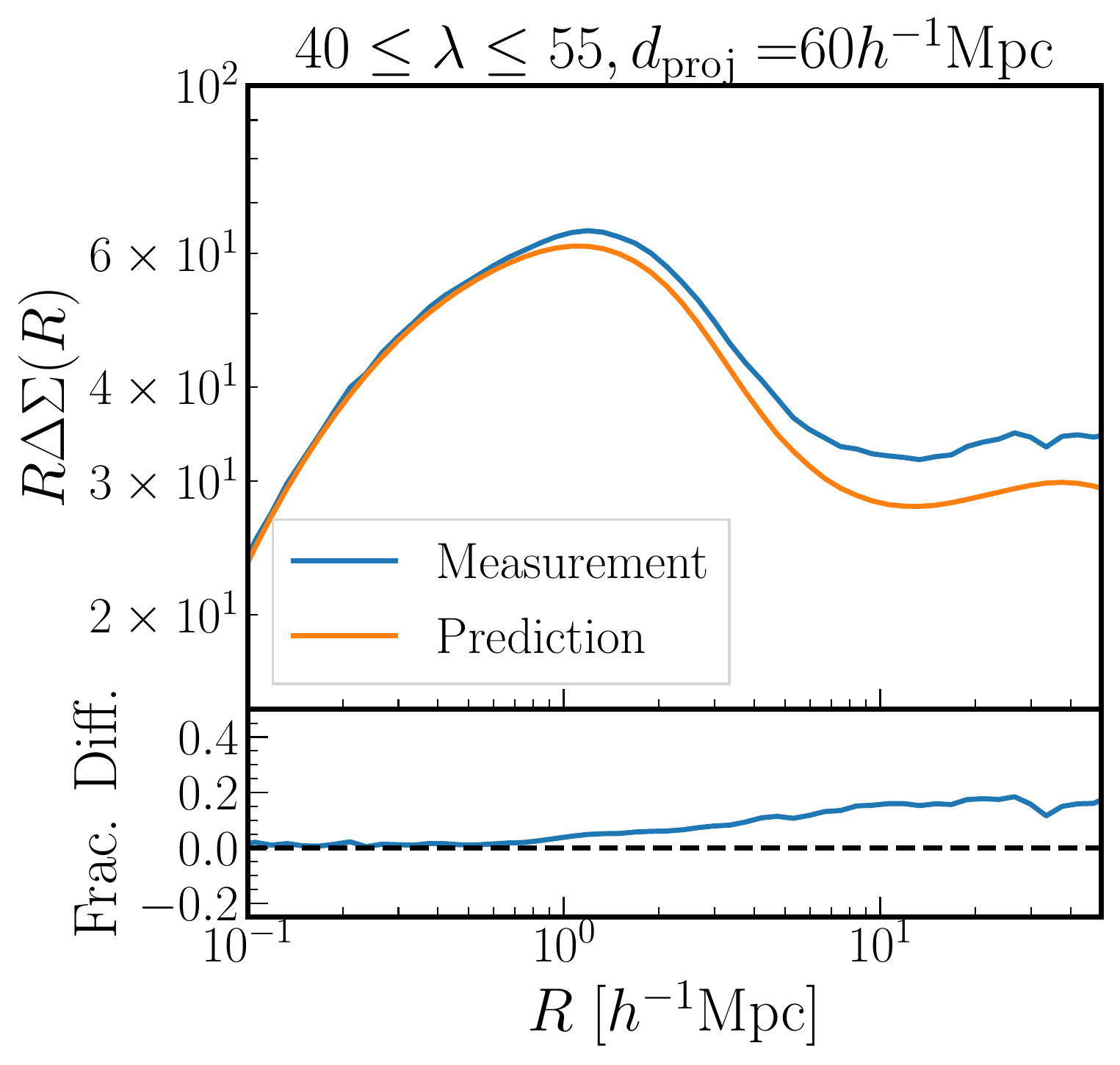}
    \includegraphics[width=0.3\textwidth]{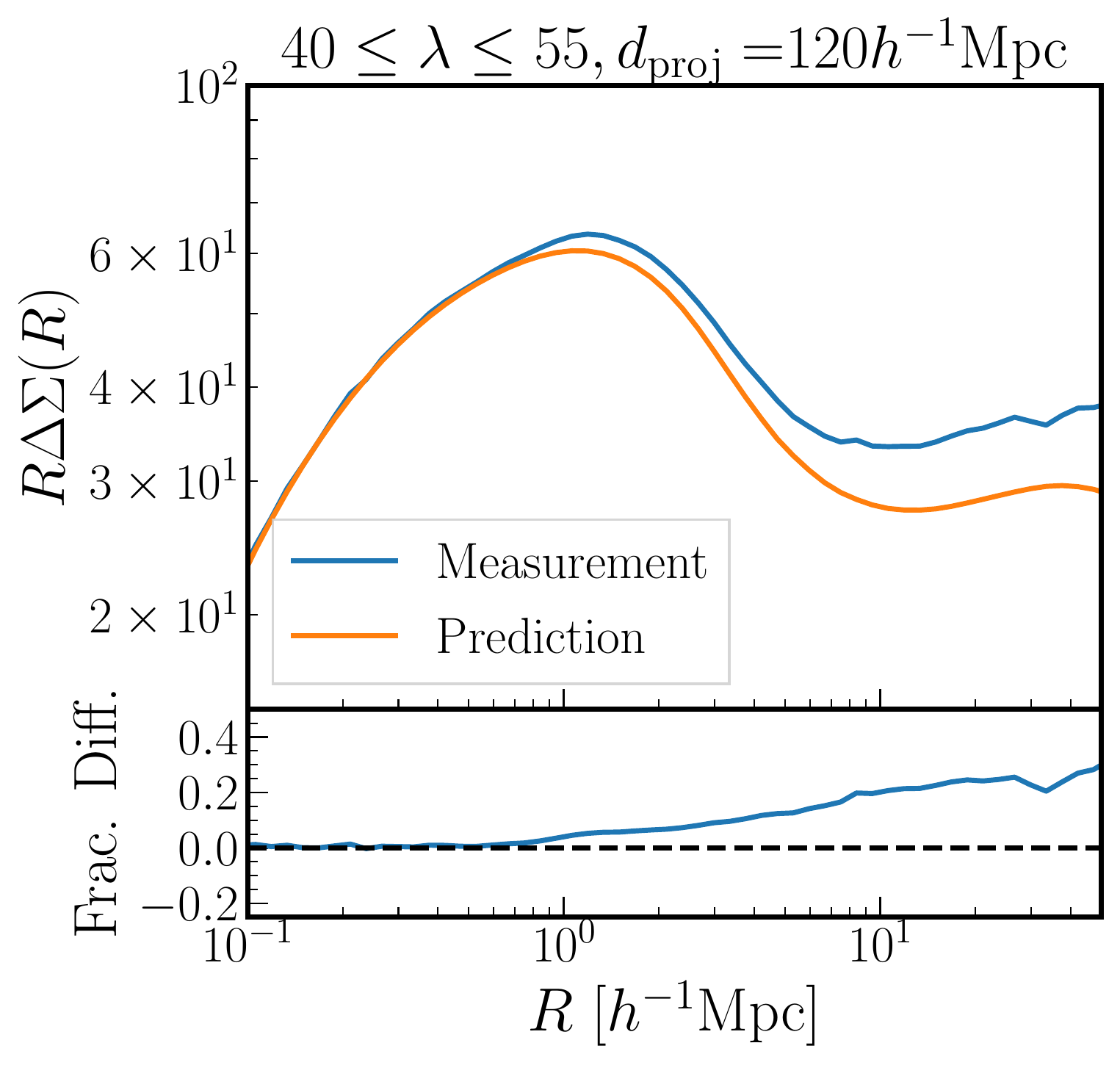}
    
    \vspace{0.3cm}
    \includegraphics[width=0.3\textwidth]{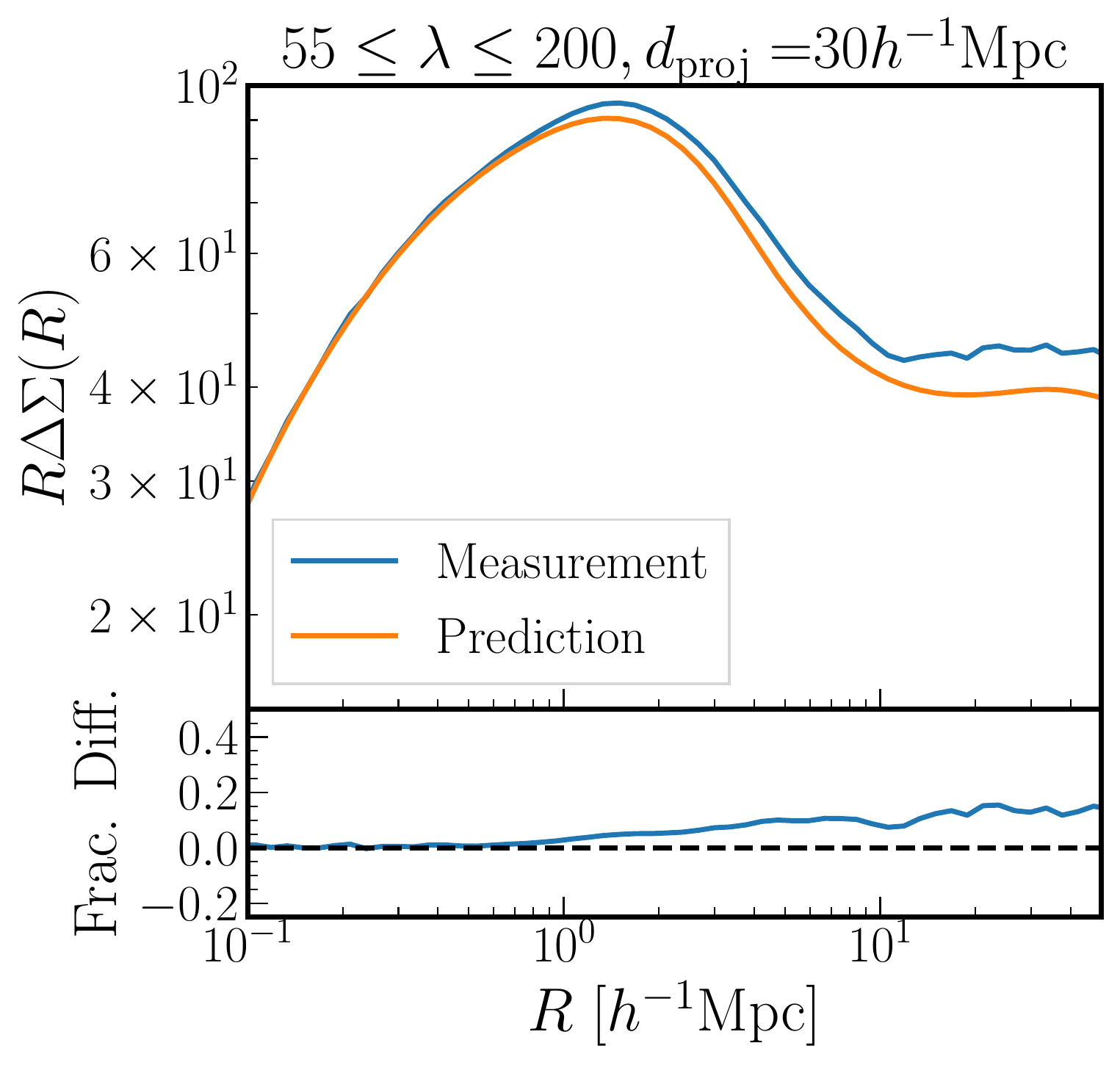}
    \includegraphics[width=0.3\textwidth]{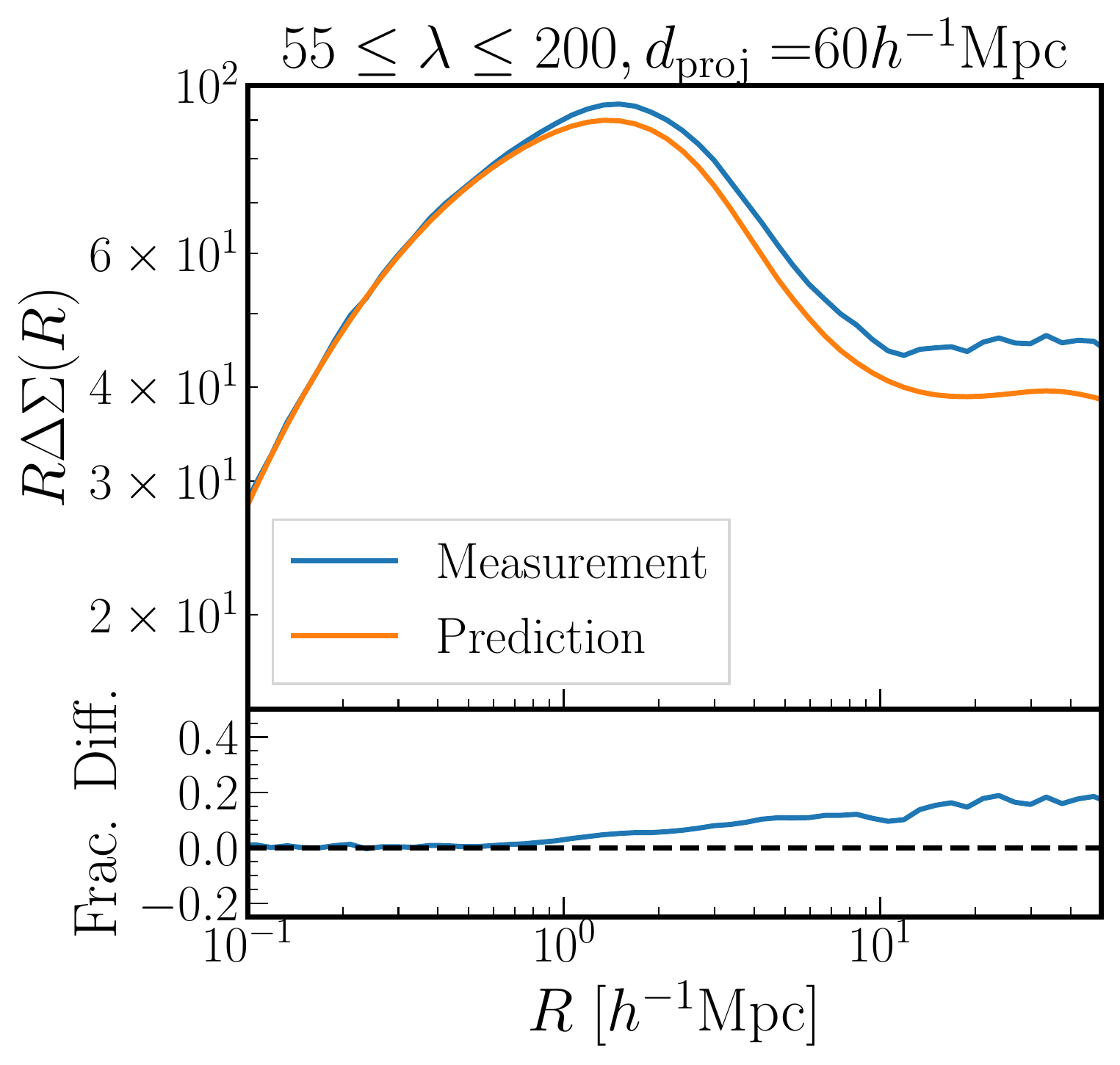}
    \includegraphics[width=0.3\textwidth]{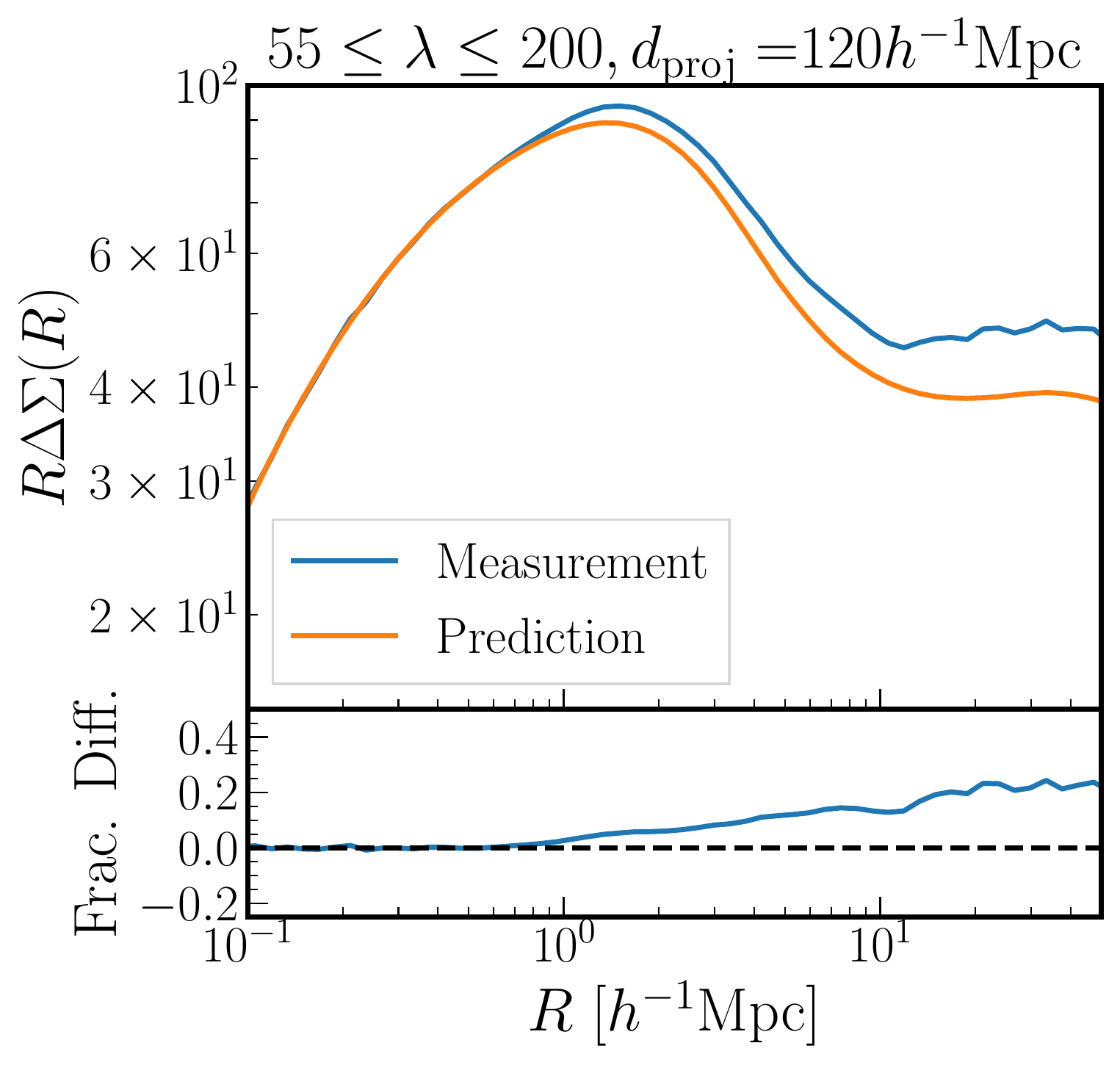}
    
    \vspace{0.3cm}

    \caption{\label{fig:all_rich1} Comparison of the measured cluster lensing signal for the mock cluster sample (blue) against the lensing signal predicted by the emulator using the primary (most massive) halo mass distribution of the sample (orange). From top to bottom, we show the results for the richness bins of $\lambda \in [20,30]$, $[30,40]$,$[40,55]$,$[55,200]$. From left to right is for the projection lengths of $d_{\rm proj}=30h^{-1}{\rm Mpc}$,$60h^{-1}{\rm Mpc}$, and $120h^{-1}{\rm Mpc}$.
        }
\end{figure*}

To validate our method, we used the values of $\alpha$ measured from the mock lensing measurements.
In \cite{Sunayama_etal2020}, the value of the boost for the lensing signals $\Delta\!\Sigma(R)$ and cluster auto-correlation functions $w_{\rm cc,obs}(R)$ were consistently proportional to $(1+\alpha)$ and $(1+\alpha)^2$ such as 

\begin{eqnarray}
\Delta\!\Sigma_{\rm obs}(R) & = & (1+\alpha)\Delta\!\Sigma_{\rm emu}(R) \\
w_{\rm cc,obs}(R) & = & (1+\alpha)^2 w_{\rm cc,emu}(R).
\end{eqnarray}
So, the measured value of $\alpha$ from the lensing signals can be used as a reference.

In this Appendix, we describe how we measured $\alpha$ from the mock lensing signals.
To measure lensing signals, we used the same cluster mock catalog and made measurements of the cluster lensing signals following \cite{Valageas_Nishimichi2011}. The details are discussed in Sec. 2.5 of \cite{Sunayama_etal2020}.
We compared the measured lensing signal $\Delta \Sigma_{\rm obs}(R)$ to the theoretical prediction $\Delta \Sigma_{\rm emu}(R)$ to measure the value of $\alpha$.
To compute theoretical predictions, we used the emulator \textit{darkemu} developed in \cite{darkemu}. 
The \textit{darkemu} takes a cosmological model, halo mass, and redshift as input parameters and makes predictions for the halo statistics assuming statistical isotropy. We make use of this isotropic prediction to compare against the measured lensing signals to isolate the boost due to the projection effects.

Fig.~\ref{fig:all_rich1} shows the lensing profiles measured from
the cluster samples against the corresponding emulator predictions. 
The emulator predictions are based on the mass of the primary (most massive) halo within the cluster region.
The figures are for all richness bins (from top to bottom) as well as $d_{\rm proj}=30h^{-1}{\rm Mpc}$, $60h^{-1}{\rm Mpc}$, and $120h^{-1}{\rm Mpc}$ (from left to right).
The fractional difference between $\Delta \Sigma_{\rm obs}(R)$ and $\Delta \Sigma_{\rm emu}(R)$ is equal to the size of the anisotropic boost $\alpha$, and the size of the boost is almost identical for all richness bins with the same $d_{\rm proj}$, while it is proportionally larger for a larger $d_{\rm proj}$.

We measured $\alpha$ using this fractional difference through the least $\chi^2$ fitting:
\begin{align}
\chi^2=\sum_{ij}(\frac{\Delta \Sigma_{\rm meas,i}}{\Delta \Sigma_{\rm pred,i}}-1-\alpha)C^{-1}_{ij}(\frac{\Delta \Sigma_{\rm meas,j}}{\Delta \Sigma_{\rm pred,j}}-1-\alpha),
\label{eq:chisq}
\end{align}
where $i,j$ are the indices of $R$ bins and $\Delta \Sigma_{\rm meas/pred,i}$ are the measured and predicted lensing profiles at $R_i$. 
The covariance matrix $C_{ij}$ is computed from the mock lensing measurements using the $(0.5h^{-1}{\rm Gpc})^3$ sub-boxes,

\begin{align}
C_{ij}=\frac{N-1}{N}\sum_{l=1}^{l=N}(\Delta\!\Sigma^{l}_{{\rm obs},i}-\left <\Delta\!\Sigma_{{\rm obs},i} \right >)(\Delta\!\Sigma^{l}_{{\rm obs},j}-\left <\Delta\!\Sigma_{{\rm obs},j} \right >),
\end{align}
where $N=136$, $l$ is the index of the jackknife samples, $i,j$ are indices of $R$ bins, and $\left <\Delta\!\Sigma_{{\rm obs},i} \right >$ is the mean from all the jackknife samples at $R_i$.
To constrain $\alpha$, we used $10h^{-1}{\rm Mpc} \leq R \leq 40h^{-1}{\rm Mpc}$.

\begin{figure}
    \centering
    \includegraphics[width=0.45\textwidth]{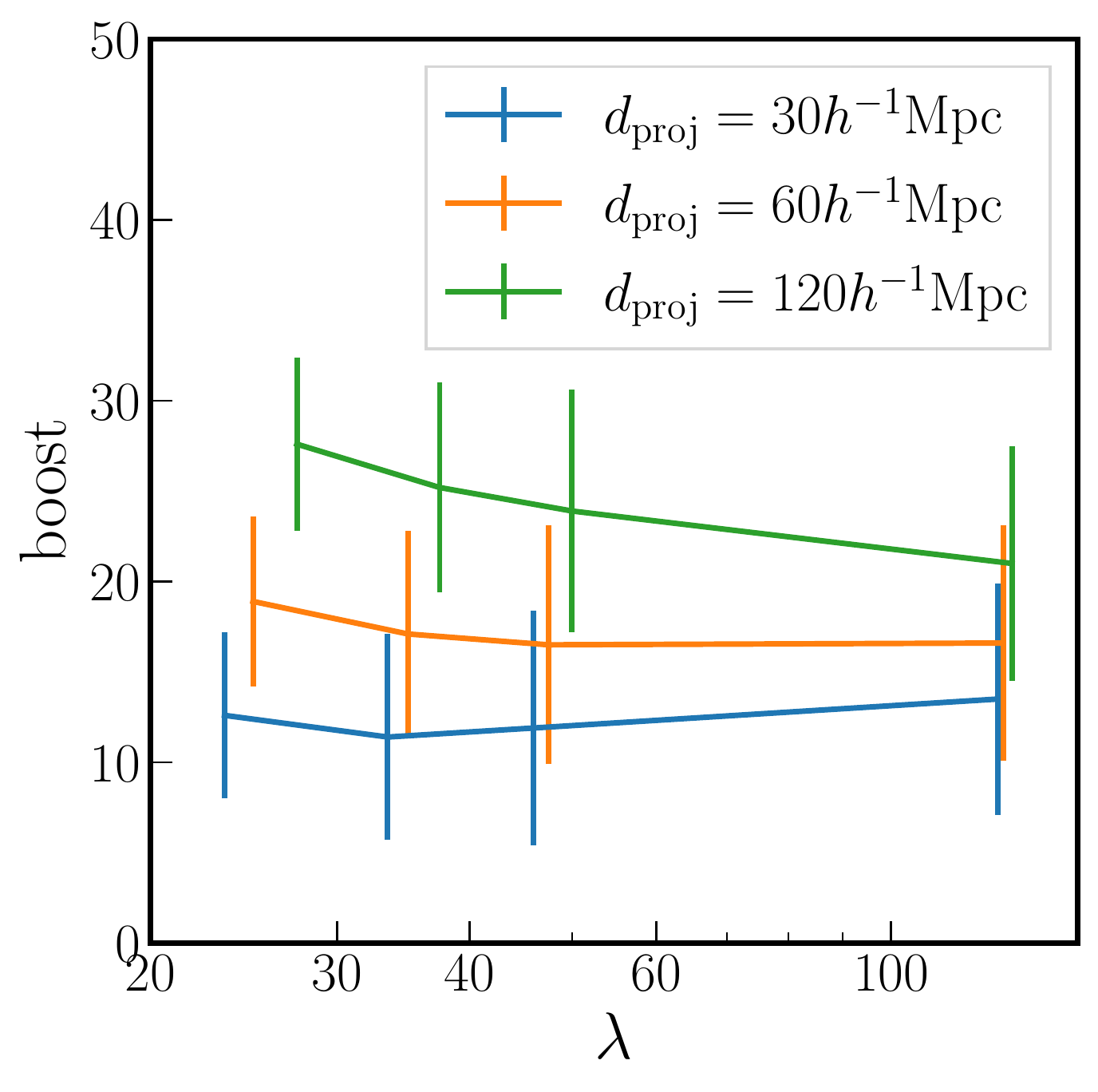}
    \caption{\label{fig:boost_richness} The best-fit values of the anisotropic boost parameter $\alpha$ as a function of richness $\lambda$ measured from the mock lensing profiles for all the cluster samples with $d_{\rm proj}=30h^{-1}{\rm Mpc}$,$60h^{-1}{\rm Mpc}$, and $120h^{-1}{\rm Mpc}$.
        }
\end{figure}

Fig.~\ref{fig:boost_richness} shows the boost $\alpha$ as a function of richness. The size of the boost shows little dependence on richness except for the cluster samples with $d_{\rm proj}=120h^{-1}{\rm Mpc}$. 
For the cluster samples with larger $d_{\rm proj}$, the size of the boost gets larger as expected.

\bsp	
\label{lastpage}
\end{document}